\documentclass[fleqn, 11pt]{wlscirep}
\usepackage[T1]{fontenc} 
\usepackage{bm}
\usepackage{chngcntr} 

\newcommand{\red}[1]{\textcolor{black}{{#1}}}  

\title{Gaussian curvature directs the distribution of spontaneous curvature on bilayer membrane necks}
\author[1]{Morgan Chabanon}
\author[1,*]{Padmini Rangamani}
\affil[1]{Department of Mechanical and Aerospace Engineering, University of California San Diego, La Jolla, CA, USA.}
\affil[*]{padmini.rangamani@eng.ucsd.edu}

\begin{abstract}
Formation of membrane necks is crucial for fission and fusion in lipid bilayers. In this work, we seek to answer the following fundamental question: what is the relationship between protein-induced spontaneous mean curvature and the Gaussian curvature at a membrane neck? Using \red{an augmented} Helfrich model for lipid bilayers \red{to include membrane-protein interaction}, we solve the shape equation on catenoids to find the field of spontaneous curvature that satisfies mechanical equilibrium of membrane necks. In this case, the shape equation reduces to a variable coefficient Helmholtz equation for spontaneous curvature, where the source term is proportional to the Gaussian curvature. We show how this latter quantity is responsible for non-uniform distribution of spontaneous curvature in minimal surfaces. We then explore the energetics of catenoids with different spontaneous curvature boundary conditions and geometric asymmetries to show how heterogeneities in spontaneous curvature distribution can couple with Gaussian curvature to result in membrane necks of different geometries.
\end{abstract}

\begin{document}
 
\maketitle

\section*{Introduction}

\begin{figure*}[ht!bp]
\center
\includegraphics[width=\textwidth]{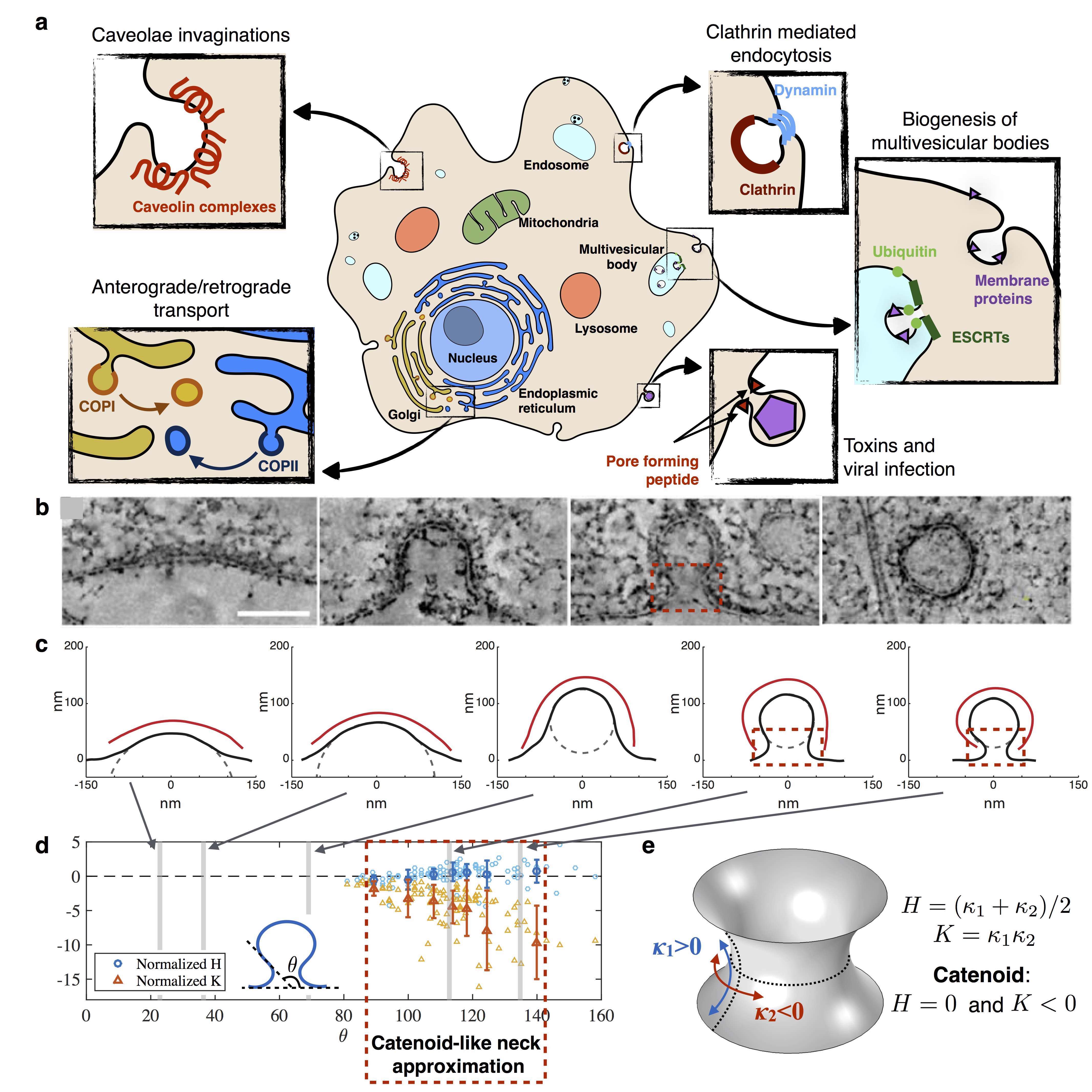}
\caption{Necks are ubiquitous in cellular membranes, and can be approximated by catenoid-like shapes. Here, we explore the relationship between catenoids and protein-induced spontaneous curvature and discuss the implications for the formation of necks. (a) The formation of necks by cellular membranes is critical for fission and fusion. A large variety of biological processes and molecular mechanisms are involved in the formation of necks; however, a common feature of these structures is that they share a catenoid-like shape. (b) Tomography slices and (c) fitted representative profiles at different stages of clathrin mediated endocytosis (CME). Red squares highlight catenoid-like neck shapes, scale bar is 100nm. (Adapted from \cite{avinoam2015}). (d) Mean ($H$) and Gaussian ($K$) curvatures (normalized with respect to neck height) at the neck of 105 CME pits, as a function of the invagination angle $\theta$. Light symbols are data points, dark symbols with error bars are average and standard deviation of 15 data points. Necks can be approximated as catenoids in the region where the mean curvature is close to zero.
(e) A catenoid belongs to the class of minimal surfaces, which have principal curvatures of opposite values at every point of the surface ($\kappa_1=-\kappa_2 $). As a consequence, the mean curvature $H=0$, and the Gaussian curvature $K<0$ everywhere.}
\label{fig:intro}
\end{figure*}


Neck-like structures are a necessary geometric intermediate for fusion and fission in cellular membranes and play important roles in membrane trafficking (both in endo- and exocytosis) and transport within the endomembrane system \cite{hurley2010, campelo2012, kukulski2012, messa2014}. Furthermore, the formation of necks is a critical step in the interaction of toxins and viral fusion proteins with cellular membranes \cite{mishra2008, schmidt2010, schmidt2013, martyna2017}. These structures are also observed in synthetic membrane systems such as in giant unilamellar vesicles subject to osmotic stress \cite{zhu2012, sanborn2013, ho2016}, lipid heterogeneities \cite{baumgart2003, baumgart2005}, protein insertion or crowding \cite{busch2015, snead2017}, and membrane-substrate interactions \cite{staykova2011}.
As shown in Fig.~\ref{fig:intro}(a), the mechanisms producing neck structures are as diverse as the underlying biological processes. Despite this diversity, their formation is subject to biophysical constraints, and most often requires bending the membrane in the presence of compositional in-plane heterogeneities. These membranes heterogeneities can be produced in many ways, resulting in a preferred mean curvature known as the spontaneous curvature \cite{zimmerberg2006, baumgart2011, lipowsky2013, jarsch2016, chabanon2017a}. The mechanisms inducing spontaneous curvature on the membrane can broadly be classified into five categories:  lipid asymmetry across the leaflets, hydrophobic insertion due to proteins, scaffolding due to proteins, oligomerization, and crowding due to proteins or other external moieties. The local value of spontaneous curvature that influences the formation of neck-like structures can then be interpreted as the combined result of several of these mechanisms. 

The formation of a neck is a complex process that can involve a large number of curvature-inducing mechanisms in sequential and parallel manners. One of the best known examples is clathrin-mediated endocytosis (CME), where a membrane coat is initiated, producing a subsequent membrane invagination and vesicle formation through the dynamic coordination of more than 30 proteins such as clathrin, epsin, BAR domain proteins and dynamin \cite{cocucci2012, kukulski2012, shi2015}. The steps involved in neck formation is illustrated in Fig.~\ref{fig:intro}(b) and (c) where tomographic slices and membrane profiles of representative steps of CME are shown (Images and data in Fig.~\ref{fig:intro}(b-d) are adapted from \cite{avinoam2015}). The stage of endocytosis can be related to $\theta$, the maximum angle between the flat plasma membrane and the bud (Fig~\ref{fig:intro}(d)) \cite{avinoam2015}. As seen from Fig~\ref{fig:intro}(b-d), no neck is yet formed at early endocytic stages corresponding to $\theta<90^\circ$. However for $\theta$ between $90^\circ$ and $160^\circ$ the neck is characterized by a mean curvature close to zero, and a negative Gaussian curvature (Fig~\ref{fig:intro}(d)). Such geometrical properties are characteristic of minimal surfaces, of which a catenoid is a non-trivial example (Fig.~\ref{fig:intro}(e)). This strongly suggests that there exists a phase during CME, and likely most necking events, where the shape of the neck is well approximated as a catenoid \cite{kozlovsky2003, jiang2008, singh2012, mcdargh2016}.


%
%

In this work, we sought to understand how the spontaneous mean curvature interacts with the Gaussian curvature of catenoid-shaped membrane necks. At the continuum scale, the energy of a lipid bilayer is commonly described by the Helfrich energy \cite{helfrich1973, ou-yang1999, steigmann2003, lipowsky2013}.  Helfrich proposed an elastic strain energy for the lipid bilayer that depends on the mean and Gaussian curvature, to capture the bending effects. Because the bilayer is being modeled as a single manifold, a penalty for the mean curvature asymmetry is imposed by means of the spontaneous mean curvature, henceforth termed just `spontaneous curvature'. As a result, one can use this framework to ask, given a spontaneous curvature field, what is the resulting shape of the membrane? Such efforts have been used successfully in \cite{seifert1991, agrawal2009a, lipowsky2013, rangamani2014, hassinger2017} among others, leading to a good understanding of the role of the mean curvature and spontaneous curvature. However, the role of the Gaussian curvature on membrane physics is trickier to grasp. In most studies of membrane mechanics, the contribution of the Gaussian curvature is not significant because of the Gauss-Bonnet theorem \cite{kreyszig1991, frankel2011}, which states that the integral of the Gaussian curvature over a surface only depends on the variations of the boundaries and topology of this surface.  While this certainly means that the Gaussian curvature will not influence the energy of a system such as a closed vesicle, this is not the case, in general, for open boundaries or necks \cite{steigmann1999}. Particularly, in the case of minimal surfaces such as catenoids where the mean curvature is zero, the Gaussian curvature is the only contribution of the curvature tensor to the energy, and therefore plays a critical role in the determination of the equilibrium state of the system.


Motivated by these considerations, we ask the following questions: Given a catenoid-shaped neck connecting two reservoirs of curvature-inducing proteins, what is the spontaneous curvature field on this surface that minimizes the Helfrich energy? How is this spontaneous curvature field influenced by the neck radius? And finally, how can we modulate the field of spontaneous curvature in order to promote the necking process? To answer these questions, we use Helfrich energy of lipid bilayers \cite{helfrich1973} \red{augmented with membrane protein energy}, and conduct simulations to identify how the Gaussian curvature and the spontaneous curvature are related in catenoid-shaped necks. 
In order to proceed, we adopt an inverse problem approach to membrane biomechanics. Traditionally, the steps involved in solving the shape equation resulting from the Helfrich model are (i) fixing the distribution of spontaneous curvature and the boundary conditions and (ii) determining the shape of the membrane from resulting mean and Gaussian curvatures. In contrast, our approach consists of (i) fixing the shape of the membrane (to a catenoid) and the boundary conditions, and (ii) solving for the resulting distribution of spontaneous curvature. We argue that once the spontaneous curvature is known, one can infer the local membrane composition from constitutive relationships for specific lipid-protein coupling. Accordingly, in all of this study, the fixed quantities are the mean and Gaussian curvatures, while the \textit{spontaneous curvature is the coordinate-dependent unknown variable} \red{through the protein density}.

\section*{Model and Methods}

\subsection*{Equilibrium model of elastic membranes with heterogeneous distribution of spontaneous curvature}

In this section, we provide a concise derivation of the generalized shape equation using a local force balance. (Additional details are provided in the Supplementary Material). This approach, initally  developed in \cite{steigmann1999, rangamani2013a, rangamani2014}, results in an equivalent model to the one obtained by variational considerations \cite{steigmann2003, agrawal2009a, agrawal2009}, but accounts for local inhomogenities on the membrane \cite{steigmann1999}. 

We first summarize the assumptions under which our model is valid:
\begin{enumerate}
\item Length-scale separation: We consider membrane deformations that are much larger than the thickness of the bilayer, allowing us to treat the membrane as a \textit{thin elastic film of negligible thickness} \cite{helfrich1973}.
\item Time scale separation: We assume that the membrane deformations are much slower than the rate at which the energy is dissipated, allowing us to consider the membrane to be at \textit{mechanical equilibrium}.
\item Incompressibility: Due to the high stretching modulus of lipid bilayers, we assume the membrane to be \textit{incompressible/inextensible}. 
\item Free energy: Based on \textit{Helfrich model} of lipid bilayers \cite{helfrich1973}, we treat the lipid membrane as an elastic manifold for which the bending energy functional depends only on the local curvatures and compositional heterogeneities.
\item Spontaneous curvature: Due to the length-scale separation, the packing heterogeneities induced by heterogeneous membrane composition (lipid types, membrane proteins) cannot be explicitly represented. Instead, the effect of the bilayer composition on the preferred membrane curvature is taken into account by the \textit{spontaneous (mean) curvature} $C(\sigma)$, which depends on the local protein density ($\sigma$).
\red{
\item Protein contribution to the free energy: We assume that the presence of membrane bound protein contributes to the free energy through an additive term $A(\sigma)$ that account for protein enthalpic and entropic effects.
}
\item Homogeneous mechanical properties: We assume the mechanical properties of the membrane to be independent of the compositional heterogeneities. A discussion of the effect of varying membrane moduli can be found in \cite{hassinger2017}.
\item Inviscid lipid membrane: Due to the fluid characteristics of lipid membranes, we neglect any resistance to shear forces in the membrane. 
\end{enumerate}

\paragraph*{Model development}

We start by expressing the local force balance on the membrane by the equation of mechanical equilibrium of an elastic surface $\omega$ subject to a lateral pressure $p$. This can be written in the compact form as \cite{steigmann1999}
\begin{equation} \label{eq:force_balance}
\bm{\sigma}_{;\alpha}^\alpha + p \mathbf{n} = \mathbf{0} \; , 
\end{equation}
where $\bm{\sigma}^\alpha$ are the stress vectors and $\mathbf{n}$ is the unit normal to the local surface. Here onwards, Greek indices range over ${1,2}$, and if repeated, are summed over this range. The semicolon indicates covariant differentiation. The stress vectors are further defined as
\begin{equation} \label{eq:Asig}
\bm{\sigma}^\alpha = \mathbf{T}^\alpha + S^\alpha\mathbf{n} \;,
\end{equation}
where $\mathbf{T}^\alpha$ are the values of the tangential stress vector field that depend on the energy per unit area of the membrane, and $S^\alpha$ is the contravariant vector field that contains the normal components of the stress vector \cite{steigmann1999, rangamani2013a}  (see Supplementary Material for details).

The most common model for the free energy density of lipid membranes is the Helfrich energy \cite{helfrich1973}. \red{We extended to account for the entropic contribution of membrane-bound proteins to the areal free-energy functional such as \cite{agrawal2011}
\begin{equation} \label{eq:Wdiff} 
W(\sigma, H,K; \theta^\alpha)=k(\theta^\alpha)[H-C(\sigma )]^{2} + k_G(\theta^\alpha) K + A(\sigma ) \;,
\end{equation}%
}
Here $H$ and $K$ are the mean and Gaussian curvatures respectively. $k(\theta^\alpha)$ and $k_G(\theta^\alpha)$ are the bending and Gaussian moduli respectively, which in the general case can be dependent on the surface-coordinate $\theta^\alpha$. $C(\sigma)$ is the spontaneous (mean) curvature, which is determined by the local membrane composition, in this case the surface density of protein $\sigma$.  While it is certainly possible to propose an explicit function of $C(\sigma)$ on the protein density as is done in the Supplementary Material, we will for now, retain its general form.
\red{
Finally, $A(\sigma)$ is the contribution of the membrane-bound proteins to the free energy and $\sigma$ is the surface density of proteins.}

The local force balance of a membrane subject to the above energy functional results in normal and tangential stress balance equations, commonly referred to as the shape equation and the incompressibility condition respectively. Following the procedure in  \cite{steigmann1999, rangamani2013a} (see Supplementary Material), these are written in the general case of anisotropic membranes as
\begin{equation} \label{eq:hape}
\Delta \left[ k (H-C) \right] + 2H\Delta k_G - (k_G)_{;\alpha\beta} b^{\alpha\beta} + 2k(H-C)(2H^2-K) + 2H(k_G K - W(\sigma,H,K) ) = p + 2\lambda H \;,
\end{equation}
and
\red{
\begin{equation}\label{eq:incomp}
\nabla \lambda = -W_\sigma \nabla \sigma
-\nabla k (H-C)^2 - \nabla k_G K \;,
\end{equation} 
where $\Delta(\cdot)$ and $\nabla (\cdot)$ are the surface Laplacian and gradient respectively, and $W_\sigma$ is the partial derivative of the free energy with respect to $\sigma$. By definition, $W_\sigma$ is the chemical potential associated the membrane proteins.}

Together, Eqs.~\ref{eq:hape} and \ref{eq:incomp} describe the equilibrium configuration of a lipid membrane subject to a distribution of protein density $\sigma$. 
Additionally, the membrane must satisfy the existence of a tangential velocity field $u^\alpha$, such that the membrane incompressibility condition $u_{;\alpha }^{\alpha }=2Hw $ is respected ($w$ is the normal velocity of the membrane). But this velocity field is decoupled from the shape equation and tangential equation for an inviscid membrane \cite{rangamani2013a}. Therefore, the sufficient condition is the existence of the flow field. In the case of static surfaces, tangential flow fields are known to exist without affecting the shape of the surface and are studied in soap films and membranes \cite{bahmani2015}.

\paragraph*{Static distribution of curvature-inducing proteins on minimal surfaces}

Eqs.~\ref{eq:hape} and \ref{eq:incomp} link the membrane geometry of the membrane (through $H$, $K$ and the surface differential operators) to the protein density ($\sigma$) and therefore the distribution of spontaneous curvature $C(\sigma)$. Here we examine how a catenoid-shaped membrane influences $C$ by specializing Eqs.~\ref{eq:hape} and \ref{eq:incomp} to minimal surfaces.
Catenoids belong the to the mathematical family of minimal surfaces, which locally minimize their surface energy everywhere \cite{osserman1996, torquato2004, frankel2011}. These have been historically studied in the context of soap films and the formation of membrane tethers \cite{kreyszig1991, powers2002, frankel2011, goldstein2010} and more recently in the context of shapes adopted by organelle membranes \cite{snapp2003, terasaki2013, bahmani2015}. 

Minimal surfaces are characterized by the property that the mean curvature
vanishes pointwise ($H=0$ everywhere on the membrane). Furthermore, as a first approximation, we consider membranes with isotropic mechanical properties ($k$ and $k_G$ are constants). Accordingly, in the absence of transmembrane pressure, the shape equation (Eq.~\ref{eq:hape}) reduces to a variable-coefficient Helmholtz equation for the spontaneous curvature
\begin{equation} \label{eq:hape_mini}
\Delta C(\sigma)-2KC(\sigma) =0 \;.
\end{equation}%
In view of the fact that the Gaussian curvature is determined entirely by the metric of the surface (Gauss' \textit{Theorema Egregium}), it follows that the spontaneous curvature on a minimal surface induces a distribution on any other such surface that can be obtained from it by an isometric map. This is the case for instance between catenoids and helicoids, provided that the boundary conditions follow the same mapping.

\red{Both the energetic contribution of the proteins $A(\sigma)$ and the local Lagrange multiplier $\lambda$ are now absent from the shape equation \ref{eq:hape_mini}, therefore uncoupling them from the incompressibility condition. Yet, any solution of Eq.~\ref{eq:hape_mini} is restricted to the condition that the balance equation \ref{eq:incomp} is satisfied. This latter condition is derived for minimal surfaces in the Supplementary Material, resulting in
\begin{equation} \label{eq:admiss_lambda}
\lambda (\sigma) =-[A(\sigma )+kC(\sigma )^{2}] + \lambda_0 \;,
\end{equation}
}
where $\lambda_0$ is a constant.

The system composed of Eqs.~\ref{eq:hape_mini} and \ref{eq:admiss_lambda}, can be further developed to explicitly depend on the local membrane composition, as shown in the Supporting Material. However, \red{explicitly defining $A(\sigma)$ and $C(\sigma)$} requires several simplifying assumptions on the type of curvature-inducing proteins, the relationship between spontaneous curvature and protein density, as well as on the protein-protein interactions. While this extra step can be conceptually insightful, it is of low practical interest in the absence of available experimental data to test this model. \red{
Furthermore, in the case of minimal surfaces, the uncoupling between the shape equation and the incompressibility condition allows us to treat the problem in term of $C(\sigma)$ without having to define the protein contribution to the membrane energy $A(\sigma)$.
} 
For this reason, we choose to solve this model in the general case and discuss the results in terms of spontaneous curvature instead of protein density. Provided the knowledge of the explicit function for $C(\sigma)$ from experiments or such as the ones proposed in the Supplementary Material, one can map $\sigma$ from $C$.

\paragraph*{Boundary conditions and relation to membrane tension}

The shape equation for minimal surfaces (Eq.~\ref{eq:hape_mini}) now has a one-way coupling with the membrane tension. Therefore we can first solve Eq.~\ref{eq:hape_mini} for $C$ given a minimal surface of Gaussian curvature $K$. From a mathematical perspective, Eq.~\ref{eq:hape_mini} is a variable-coefficient Helmholtz equation for $C$, and two boundary conditions either Dirichlet, Neumann, or of the mixed type \cite{rangamani2013a} can be prescribed. In this work, we consider Dirichlet conditions at the minimal surface boundaries. The physical justification for this choice is that necks are connecting larger membrane reservoirs on each sides, which serve as sources for curvature-inducing proteins (see Fig.~\ref{fig:cat1}(a)). Since the distribution of $C$ on a minimal surface should satisfy the admissibility conditions for $\lambda$ (Eqs.~\ref{eq:admiss_lambda}), imposing Dirichlet boundary conditions for $C$ results in imposing the membrane tension $\lambda$ at the boundary satisfying Eq.~\ref{eq:admiss_lambda}. In fact, this condition provides a correspondence between spontaneous curvature and membrane tension everywhere on the surface. 

It is important to note that the boundary conditions here should not be confused with boundary conditions between the protein inclusion and the monolayers, which have been studied in detail elsewhere \cite{aranda-espinoza1996, kim1998, gil1998}. In contrast to these studies, here we consider the spontaneous curvature of the whole bilayer, including the one induced by the protein inclusion. In this approach, we neglect any membrane thickness variation (see model assumption 1), and do not account for the specific location of the protein inclusion. The only boundaries we consider are the ones at the boundaries between the membrane minimal surface and the membrane reservoirs.

\subsection*{Model Implementation}
%

\paragraph*{Catenoids}

We consider a catenoid of height $h_0$ and neck radius $r_n$ such as the one depicted in Fig.~\ref{fig:cat1}(b). In axisymmetric coordinates, this surface can be parametrized by 
\begin{equation}
r = r_n\cosh(z/r_n) \quad \text{with } z\in[-h_0/2; h_0/2] \;.
\end{equation}
We seek the distribution of spontaneous curvature along the arclength $s = r_n \sinh(z/r_n)$ in the axial direction. We choose the total arclength $L = 2r_n \sinh[h_0/(2r_n)]$, as the characteristic length of the catenoid.

The shape equation (Eq.~\ref{eq:hape}) involves two geometrical invariants of the surface: the mean and the Gaussian curvature. The mean curvature is zero everywhere on a catenoid; however the Gaussian curvature of a catenoid depends on $z$ and the neck radius as
\begin{equation} \label{eq:K}
K = -\left[\frac{1}{r_n \cosh^2(z/r_n)}\right]^2 = -\left[\frac{1}{r_n (1+(s/r_n)^2) } \right]^2.
\end{equation}
The Gaussian curvature of the catenoid is negative everywhere and is minimum (i.e. maximum magnitude) when $z=0$ or $s=0$ (Figs.~\ref{fig:cat1}(c) and \ref{fig:cat2}(b)). As the neck radius decreases, the Gaussian curvature at the neck decreases towards minus infinity, while it tends to zero away from the neck. 

\paragraph*{Boundary conditions}

As noted above, we specify the spontaneous curvature at the boundaries with the following Dirichlet boundary conditions:
\begin{equation} \label{eq:bc}
C= \left\{
\begin{array}{l}
C_0 \quad \text{at the lower boundary} \\
C_1 \quad \text{at the upper boundary}
\end{array}\right. ,
\end{equation}
where $C_0$ and $C_1$ are prescribed. 

\paragraph*{Implementation}

We write the dimensionless forms of Eqs.~\ref{eq:hape_mini} and \ref{eq:bc} using the total arclength of the symmetric catenoid $L$, and the reference value of spontaneous curvature at one of the boundaries $C_0$, respectively.  Accordingly, the geometric variables are scaled as $\bar{\theta}^\alpha = \theta^\alpha/L$, and $\bar{K} = KL^2$, while the spontaneous curvature is scaled as $\bar{C} = C/C_0$. Accordingly, the system can be written in its dimensionless form as
\begin{equation} \label{eq:hape_mini_nd}
\Delta \bar{C} -2\bar{K}\bar{C} =0,
\end{equation}
with the dimensionless boundary condition
\begin{equation} \label{eq:bc_nd}
\bar{C}= \left\{
\begin{array}{l}
1 \quad \text{at the lower boundary} \\
C_1/C_0 \quad \text{at the upper boundary}
\end{array}\right. \;.
\end{equation}

The bending energy of the membrane is defined by the integral of the contribution from the curvature to the energy density over the catenoid surface ($\Omega$), that is
\begin{equation}
W_B = \int \left(k C^2 + k_GK \right) d\omega \;.
\end{equation}
In dimensionless form, the total bending energy is written
\begin{equation}
\frac{W_B}{k C_0^2 L^2} = \int \left( \bar{C}^2 + \frac{k_G}{k}\frac{1}{C_0^2L^2}\bar{K} \right) d\bar{\omega} \;.
\end{equation}

Dividing by the dimensionless area of the catenoid $\bar{\Omega}=\Omega/L^2$, the dimensionless energy per area is $W_B/(k C_0^2 \Omega)$. We use the value $k_G/k=-0.9$ for the ratio of the Gaussian to bending modulus \cite{hu2012}.

To our knowledge, no analytical solution for a variable coefficient Helmholtz equation of the form of Eq.~\ref{eq:hape_mini_nd} is available. Therefore, we solve the system composed of Eqs.~\ref{eq:hape_mini_nd} and \ref{eq:bc_nd} with the finite element solver COMSOL Multiphysics\textsuperscript{\tiny\textregistered} 5.2a, and the `Surface reaction' module that has a built-in surface Laplacian. Parametric studies are conducted by exploiting the COMSOL model with the `COMSOL with Matlab' module.

\section*{Results}


\paragraph*{The Gaussian curvature of the catenoid governs the distribution of spontaneous curvature.}

\begin{figure*}[t!bp]
\center
\includegraphics[width=\textwidth]{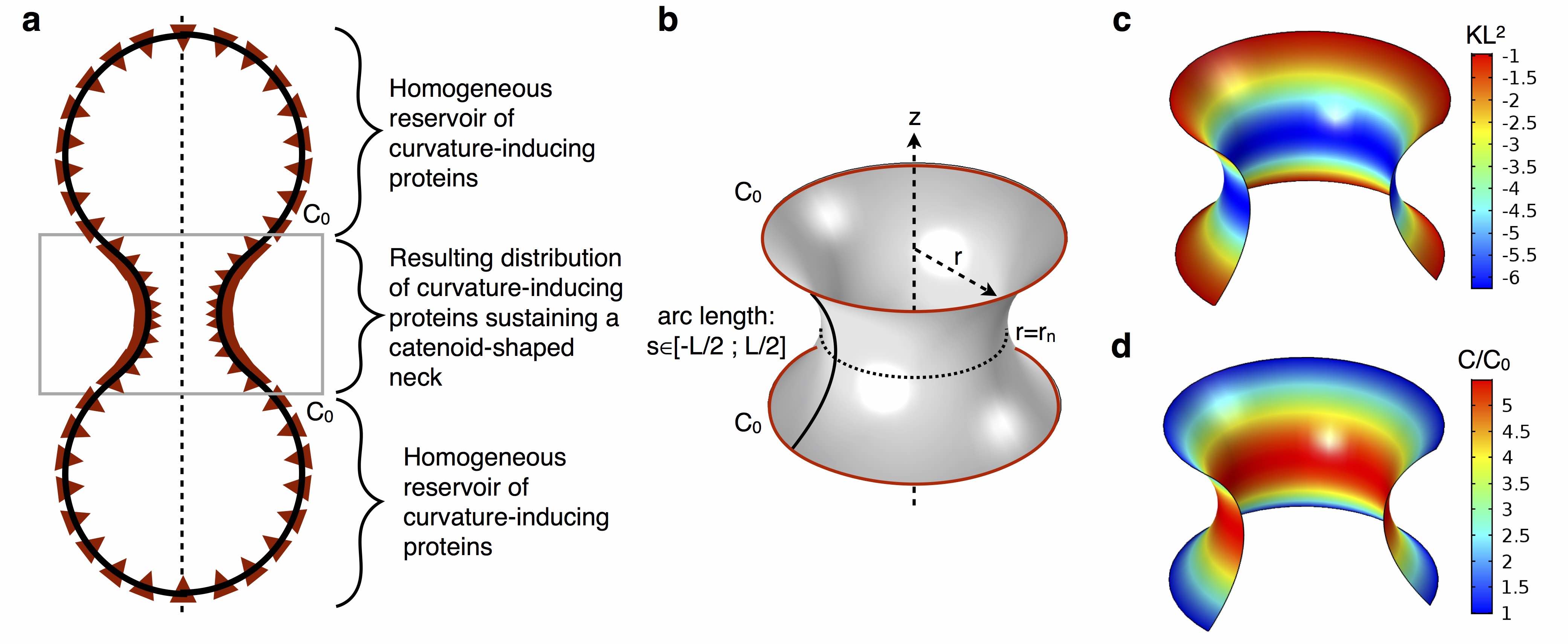}
\caption{Given a neck radius $r_n$, a catenoid-shaped neck connected to two identical reservoirs of curvature-inducing proteins shows a variation of spontaneous curvature along the arclength. (a) Schematic of a possible spontaneous curvature-inducing protein distributed along a catenoid connecting two vesicles. The two spherical parts are protein reservoirs imposing spontaneous curvature at the boundaries of the catenoid, and resulting in a distribution of curvature-inducing proteins along the neck.
(b) Geometry and boundary conditions of the catenoid. The arclength $s$ varies from $-L/2$ to $L/2$ and the boundary conditions are $C=C_0$. (c) Variation of dimensionless Gaussian curvature $K/L^2$ on a catenoid of neck radius $r_n=0.4L$. (d) Corresponding distribution of dimensionless spontaneous curvature. }
\label{fig:cat1}
\end{figure*}

\begin{figure*}[t!bp]
\center
\includegraphics[width=\textwidth]{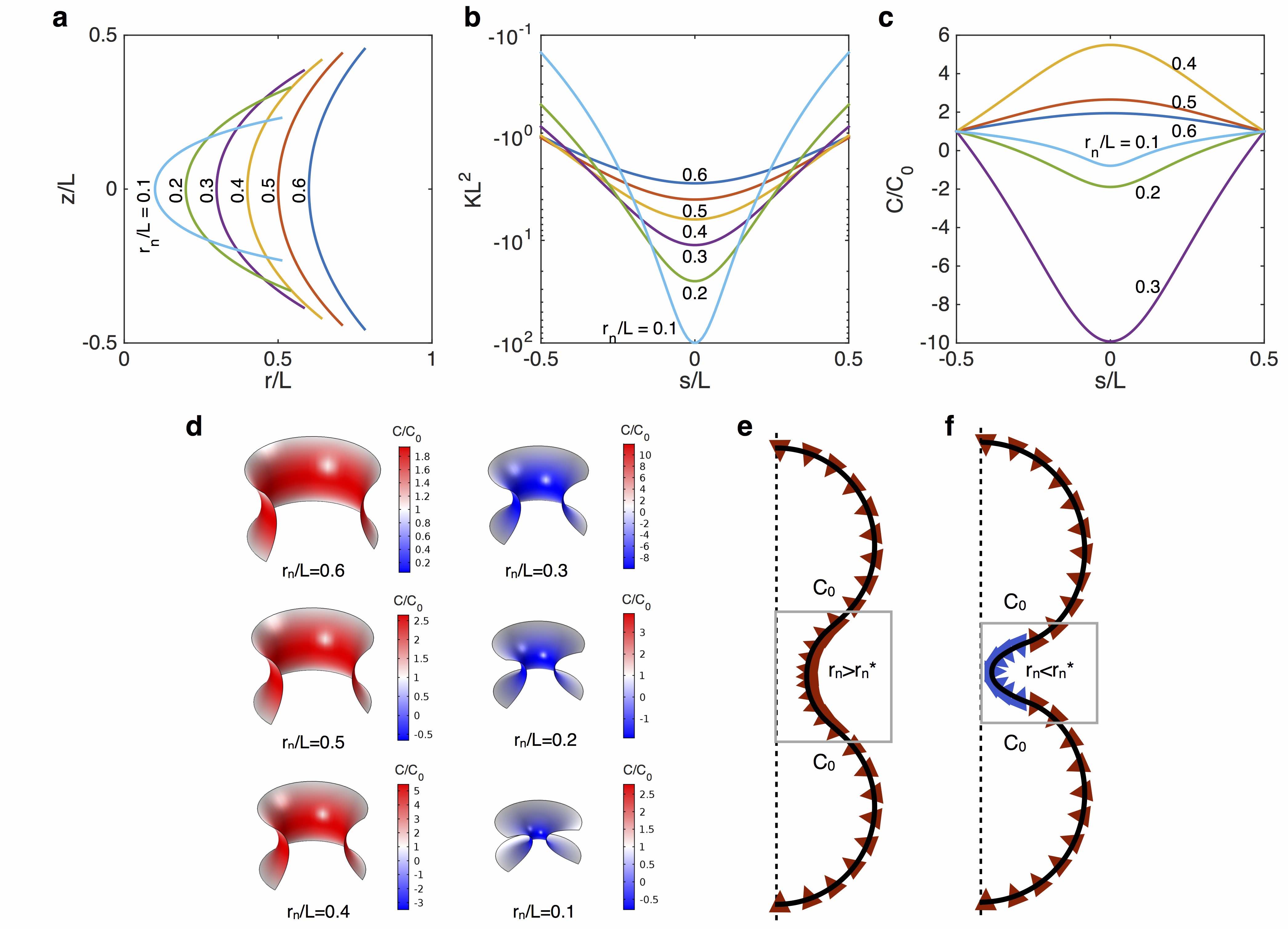}
\caption{Constraining the neck of a catenoid induces a switch in the sign of spontaneous curvature at constant total arclength and boundary conditions. (a) Axisymmetric geometry of six catenoids of various neck radii $r_n$ and same total arclength $L$. (b) Dimensionless Gaussian curvature along the arclength $s$ for various neck radii. (c) Resulting spontaneous curvature along the arclength for $C=C_0$ at both boundaries. (d) Corresponding distribution of spontaneous curvature on the catenoid. (e, f) Schematics of a distribution of curvature inducing proteins on a catenoid connecting two vesicles. (e) For a catenoid-shaped neck of radius larger than the critical radius $r_n>r_n^*$, only proteins inducing a spontaneous curvature of the same sign as $C_0$ are required to sustain the catenoid. (f) When the neck radius is below the critical neck radius $r_n<r_n^*$, proteins with spontaneous curvature of opposite sign as $C_0$ are necessary to sustain the catenoid-shaped neck.}
\label{fig:cat2}
\end{figure*}

We begin our analysis with the catenoid shown in Fig.~\ref{fig:cat1}(b) with a total arclength $L$ and neck radius $r_n=0.4L$. The geometry of a catenoid determines its Gaussian curvature along its arclength through Eq.~\ref{eq:K}. The Gaussian curvature of the catenoid considered here is displayed in Fig.~\ref{fig:cat1}(c). As expected $K$ is negative everywhere, with a maximum magnitude at the neck. How does the Gaussian curvature affect the distribution of protein-induced spontaneous curvature? We answer this question by solving the boundary value problem composed of Eq.~\ref{eq:hape_mini_nd} with $C=C_0$ at both boundaries (see Fig.~\ref{fig:cat1}(b)). The resulting field of spontaneous curvature is shown in Fig.~\ref{fig:cat1}(d). For this configuration, the spontaneous curvature is positive everywhere with a maximum at the neck, following the intensity of Gaussian curvature.

In order to interpret this spatially varying spontaneous curvature, one can think of the inclusion of a single type of conical protein into the lipid bilayer. In this case, a catenoid with fixed spontaneous curvature at the boundaries can be schematically represented as in Fig.~\ref{fig:cat1}(a), where two reservoirs of the curvature-inducing protein are connected to the boundaries. A variation of spontaneous curvature along the catenoid corresponds to a variation of protein surface density along the neck. The value of $C$ as a function of protein density depends on the protein and on the curvature inducing mechanism  \cite{lipowsky2013}.

\paragraph*{The energy required to maintain a catenoid-shaped membrane through spontaneous curvature presents a barrier at a critical neck radius.}

\begin{figure*}[t!bp]
\center
\includegraphics[width=\textwidth]{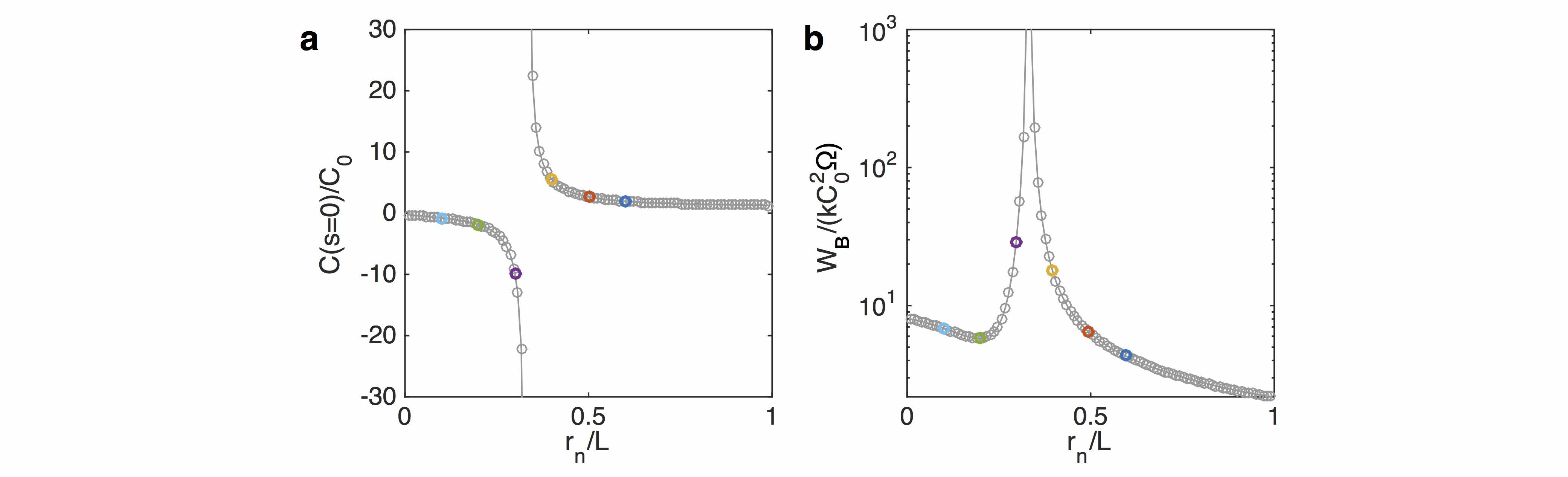}
\caption{The switch of spontaneous curvature at the neck is accompanied by an energy barrier of the catenoid. (a) Normalized membrane spontaneous curvature at the neck as a function of the neck radius. (b) Energy per area of a catenoid as a function of the dimensionless neck radius. (Color symbols correspond to neck radii configurations shown in Fig.~\ref{fig:cat2}(a-c). Solid line is a visual guide.).}
\label{fig:cat3}
\end{figure*}

The radius of the neck is an important geometric parameter of catenoids, and is of particular interest to pre-fusion events in trafficking \cite{hurley2010, campelo2012, kukulski2012, messa2014}. We next investigate how the neck radius influences the distribution of spontaneous curvature along a catenoid.  We vary the neck radius between $r_n=0.1L$ and $r_n=0.6L$ as shown in Fig.~\ref{fig:cat2}(a). The resulting Gaussian curvatures along the arclength, presented in Fig.~\ref{fig:cat2}(b), show an increase of the curvature intensity at the neck as the neck decreases. Away from the neck however, a smaller neck radius produces a lower Gaussian curvature.

We solve the boundary problem for the spontaneous curvature, resulting in the distribution of $C$ depicted in Fig.~\ref{fig:cat2}(c). The boundary conditions are the same as the one shown in Fig.~\ref{fig:cat1}(b), with $C=C_0$ at both boundaries, corresponding to a neck connected to two equal reservoirs of curvature inducing proteins. Although the maximum of spontaneous curvature intensity is always at the neck, its value is a non-monotonic function of the neck radius and Gaussian curvature. For large neck radii, decreasing $r_n$ increases the maximum of C, until a critical neck radius $r_n^*$ below which $C$ at the neck switches signs. After this switch, further decreasing the neck radius lowers the intensity of spontaneous curvature. Fig.~\ref{fig:cat2}(d) shows the distribution of spontaneous curvature on the catenoids for the corresponding radii. 

This non-intuitive switch-like behavior is surprising because neither the neck radius nor the resulting Gaussian curvature show a discontinuity. Furthermore, the boundary conditions for the spontaneous curvature are constant $C=C_0$. An intuitive understanding of this behavior can be obtained by considering a simplified case where the Gaussian curvature $K$ is constant along the arclength, reducing the shape equation for minimal surfaces to a one-dimensional simple harmonic oscillator. As shown in the Appendix, in this simplified case, reducing the neck radius is equivalent to decrease the period of the oscillator, which, when subject to fixed non-zero boundary conditions, leads to a series of diverging values for $r_n^*/L \sim \sqrt{2} / [\pi(1+2n)]$, with $n\in\mathbb{Z}$. For $n=0$ we obtain $r_n^*/L \simeq 0.45$, close to the value observed in Fig.~\ref{fig:cat3} for the catenoid. Interestingly, the value of neck radius at which the switch occurs does not depends on the value of the boundary condition, but only on the total arclength. Note that other critical values are expected for $n\neq 0$, and although these are observed for a constant $K$ (see Appendix), in the case of the catenoid, $K$ tends to zero away from the neck, suppressing the other possible switches. 

The switch in sign of the spontaneous curvature as a function of neck radius can be interpreted as a requirement that another set of proteins with spontaneous curvature opposite to the one in the reservoir will be needed to minimize the energy of catenoids with smaller necks. This idea is shown in Figs.~\ref{fig:cat2}(e) and (f), where possible distributions of curvature-inducing proteins are depicted for necks larger and smaller than the critical radius. 

To further identify the relationship between the switch in spontaneous curvature and the geometry, we plot in Fig.~\ref{fig:cat3}(a) the spontaneous curvature at the neck as a function of $r_n$. These results confirm the switch-like behavior described above and in the vicinity of a critical neck radius $r_n^*\simeq 0.33L$, the spontaneous curvature at the neck diverges, with positive values above $r_n^*$, and negative values below. In the two limits of large and small neck radii, the spontaneous curvature tends to $C_0$ everywhere. This is consistent with the two limit shapes of a catenoid: a tube and two inverted cones, both of which have a zero Gaussian curvature, and therefore no spatial variation of spontaneous curvature.

As show in Fig.~\ref{fig:cat3}(b), the bending energy required for the mechanical equilibrium of a catenoid-shaped neck through spontaneous curvature also shows an energy barrier at the critical neck radius $r_n^*$ corresponding to the switch in $C$. Away from this energy barrier, the radius of the neck can be reduced by small increases in the elastic energy of the system. The passage from one side to the other of the energy barrier will require additional mechanisms such as relaxation of the boundary conditions, external forces (e.g. actin pulling in clathrin mediated endocytosis \cite{boulant2011}), or a transient geometrical deviation from a symmetric catenoid. In the following, we investigate how the energy barrier in the catenoid-shaped neck can be modulated by spontaneous curvature and geometry.

\paragraph*{The differential in spontaneous curvature between the boundaries modulates the intensity of the energy barrier.}

\begin{figure*}[t!bp]
\center
\includegraphics[width=\textwidth]{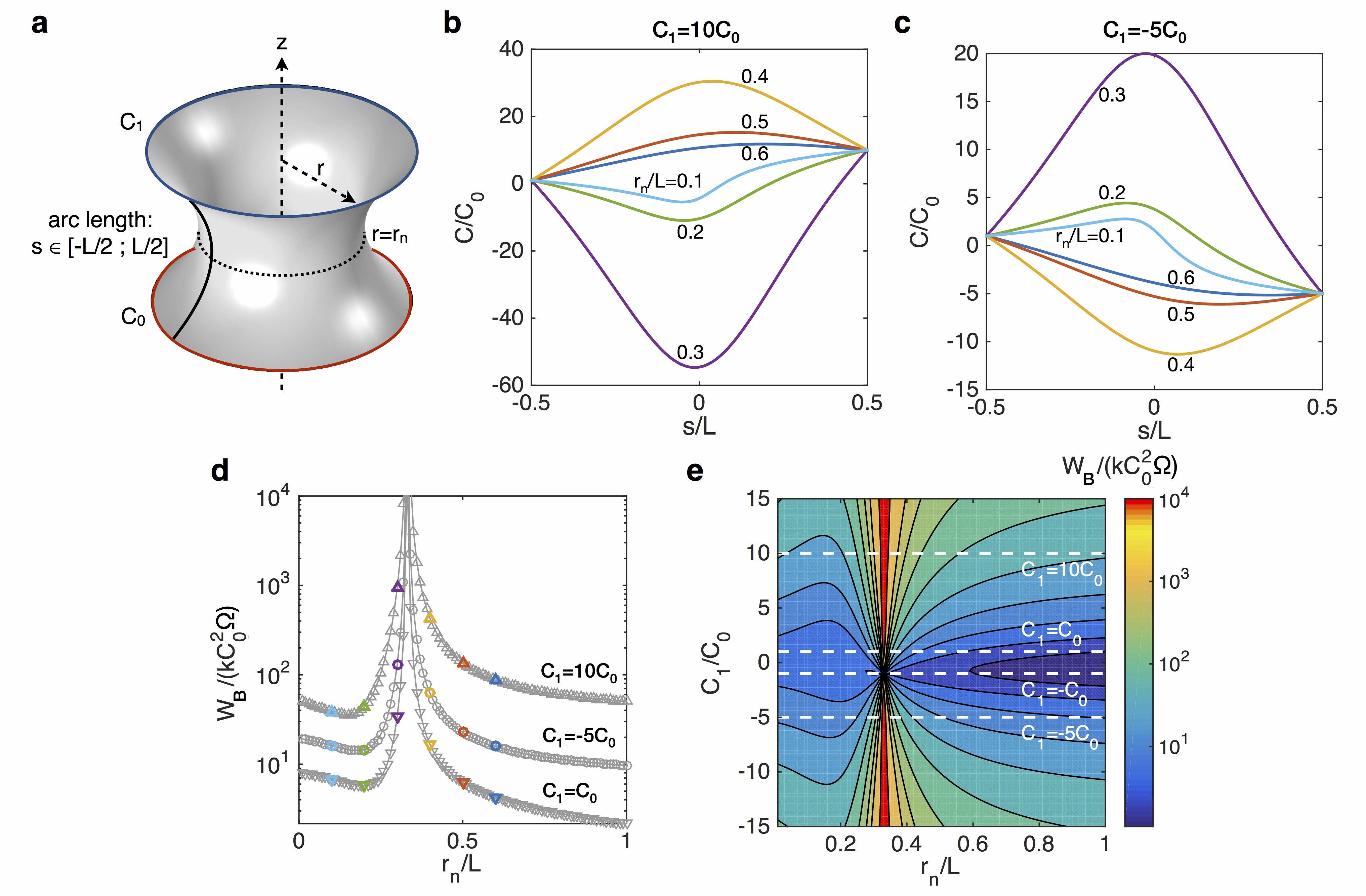}
\caption{A spontaneous curvature differential between the two boundaries determines the width of the associated energy barrier but does not influence the position of the switch in spontaneous curvature. (a) Schematic of a catenoid with unequal boundary conditions. (b) Distribution of spontaneous curvature along the arclength for $C_1/C_0=10$ for various neck radii. (c) Distribution of spontaneous curvature along the arclength for $C_1/C_0=-5$. (d) Comparison of the energy per area of the catenoid for various boundary conditions. The larger the spontaneous curvature differential at the boundary, the larger the energy barrier corresponding to the sign switch. (e) Contour plot of the energy per area for $C_1/C_0$ varying between -15 and 15 as a function of the neck radius. The energy barrier (warm colors) is located around a constant critical neck radius of about $r_n^*=0.33L$.}
\label{fig:cat_DC1}
\end{figure*}

\begin{figure*}[t!bp]
\center
\includegraphics[width=\textwidth]{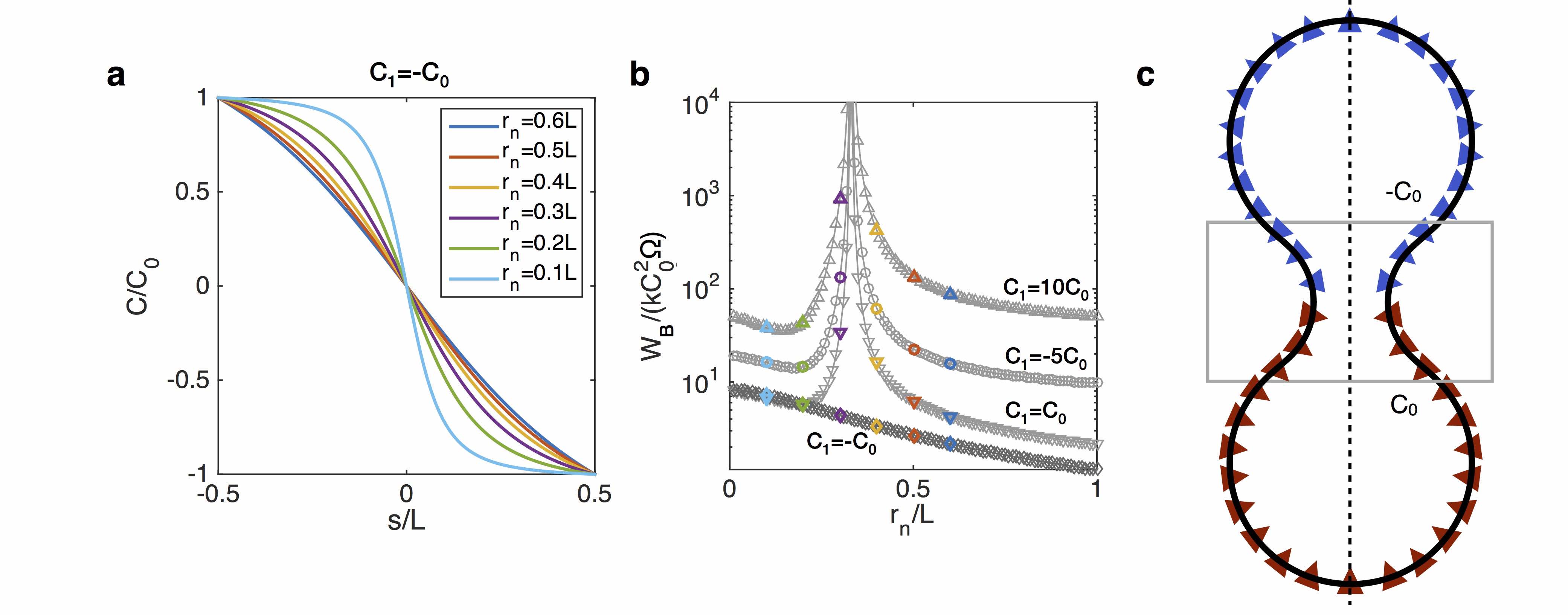}
\caption{For boundary conditions where the spontaneous curvature have opposite values ($C_1=-C_0$), the switch in the sign associated with the neck radius is nullified, and the spontaneous curvature at the neck is zero. (a) Spontaneous curvature along the arclength of a catenoid with $C_1=-C_0$ for various neck radii. (b) Energy per area of the catenoid as function of the neck radius; no energy barrier is observed for $C_1=-C_0$. (c) Schematic of a catenoid connecting two vesicles with curvature inducing proteins of opposite signs at the two spherical reservoirs. The smooth transition from $C_0$ to $-C_0$ requires a zero spontaneous curvature at the neck.}
\label{fig:cat_DC2}
\end{figure*}

How do the boundary conditions influence the distribution of spontaneous curvature in the neck? We consider the catenoid represented in Fig.~\ref{fig:cat_DC1}(a), where the spontaneous curvature at the upper boundary is $C_1\neq C_0$. This situation is likely to occur in a cellular context where the heterogeneous membrane composition may produce a differential in spontaneous curvature across the neck. We find that the switch-like behavior in spontaneous curvature persists independently of the ratio $C_1/C_0$. As shown in Figs.~\ref{fig:cat_DC1}(b) and (c), the distribution of $C$ is tilted to accommodate the boundary conditions, and the sign of the extremum is determined by the boundary condition with the largest absolute value. However the critical radius at which the switch occurs remain the same. This is better seen in Fig.~\ref{fig:cat_DC1}(d), where the critical radius associated with the energy barrier is independent of $C_1$. This observation is consistent with the expression for the critical neck obtained with the simple oscillator analogy that is independent of the boundary conditions (see Eq.~S2.6).

To fully explore the influence of the spontaneous curvature differential at the boundaries, we computed the energy of the catenoid for a wide range of $C_1/C_0$ as a function of $r_n$ (Fig.~\ref{fig:cat_DC1}(e)). The results confirm the behavior described above, except for $C_1=-C_0$ where a singularity seems to occur. The spontaneous curvature profile in this case, where the boundary of the catenoid have opposite curvatures, is shown in Fig.~\ref{fig:cat_DC2}(a). Here the switch in spontaneous curvature is suppressed, and $C=0$ at the neck independent of the neck radius. Correspondingly, the energy barrier vanishes as seen in Fig.~\ref{fig:cat_DC2}(b). Once again, this scenario can be interpreted as a neck connecting two reservoirs of proteins with the same magnitude of curvature but in opposite directions (see Fig.~\ref{fig:cat_DC2}(c)). The boundary conditions produce smooth transition from $C_1$ to $C_0$ along the catenoid, transiting by $C=0$ at the neck. This result is evident from the simple oscillator analogy, where the spontaneous curvature at the neck is proportional to $C_0+C_1$ (see Eq.~S2.4), and is therefore invariably zero for $C_1=-C_0$.

Interestingly, the energy per unit area away from the barrier suggests that for a large spontaneous differential at the boundaries, a small neck radius is energetically favorable compared to a large neck radius (Fig.~\ref{fig:cat_DC2}(b)). This result contrasts with the case $C_1=C_0$, where large neck radii are favorable (Fig.~\ref{fig:cat3}(b)).
The variation of spontaneous curvature differential across the neck may be a mechanism cells utilize to modulate the energy to form a neck. In particular, by accessing the large heterogeneity available due to membrane lipid composition and proteins, cells can disrupt the energy barrier associated with the transition from a large to a small neck radius, and \textit{vice versa}.

\paragraph*{The catenoid geometrical asymmetry modulates the location of the energy barrier.}

\begin{figure*}[t!bp]
\center
\includegraphics[width=\textwidth]{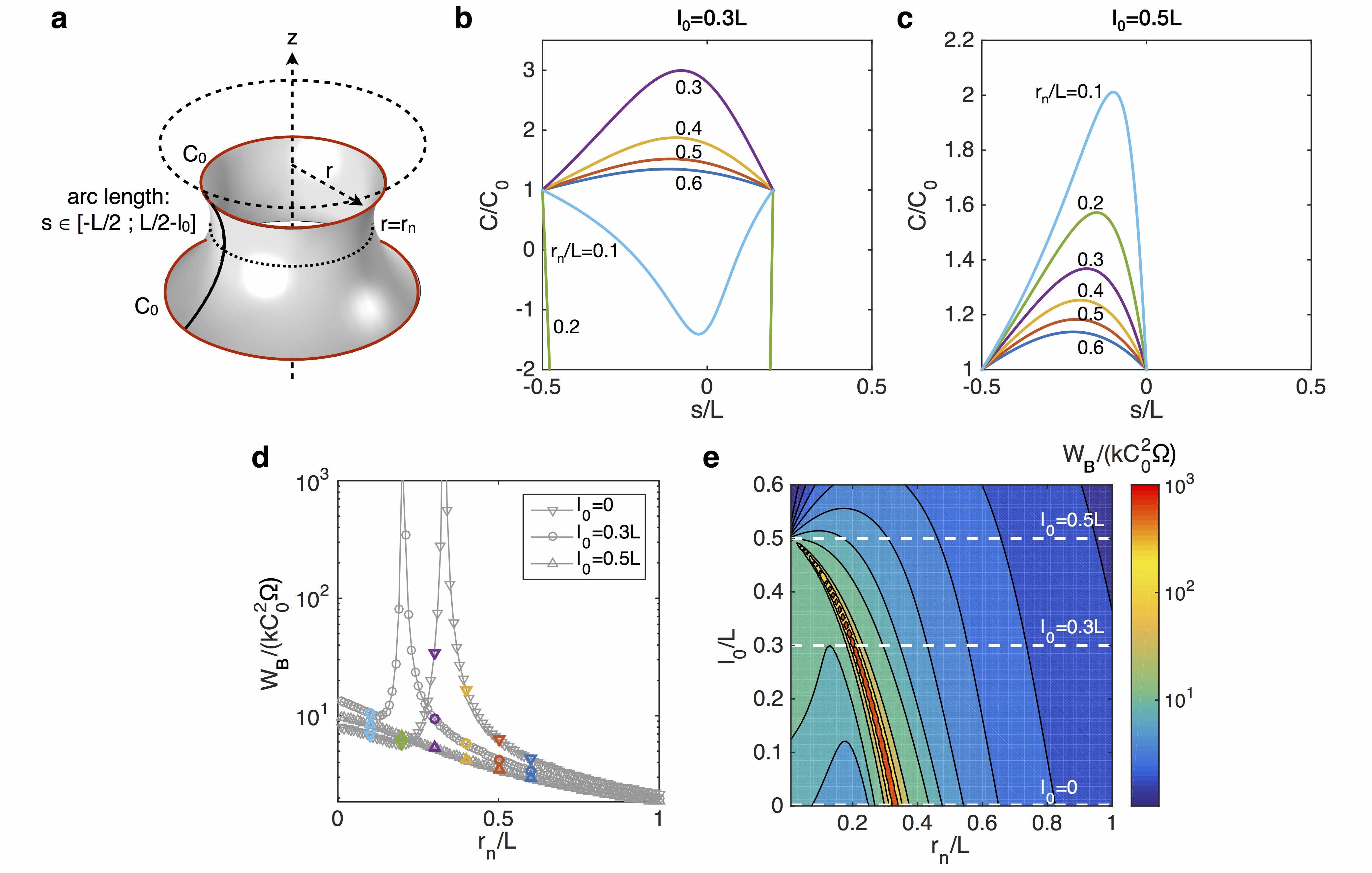}
\caption{Asymmetry along the catenoid arclength determines the value of neck radius at which the spontaneous curvature sign switches. (a) Schematic of the geometry of the catenoid with a truncated arclength $L-l_0$. (b, c) Distribution of spontaneous curvature along the arclength for $l_0/L=$0.3 (b) and $l_0/L=$0.5 (c), for various neck radii. (d) Comparison of the energy per area of the catenoid for various degrees of asymmetry. The more truncated, the smaller value of the critical neck radius. For a half catenoid ($l_0/L=0.5$), the energy barrier and the sign switch disappear. (e) Contour plot of the energy per area for $l_0/L$ varying between 0.1 and 0.6 as a function of the neck radius. }
\label{fig:cat_l}
\end{figure*}

\begin{figure*}[t!bp]
\center
\includegraphics[width=\textwidth]{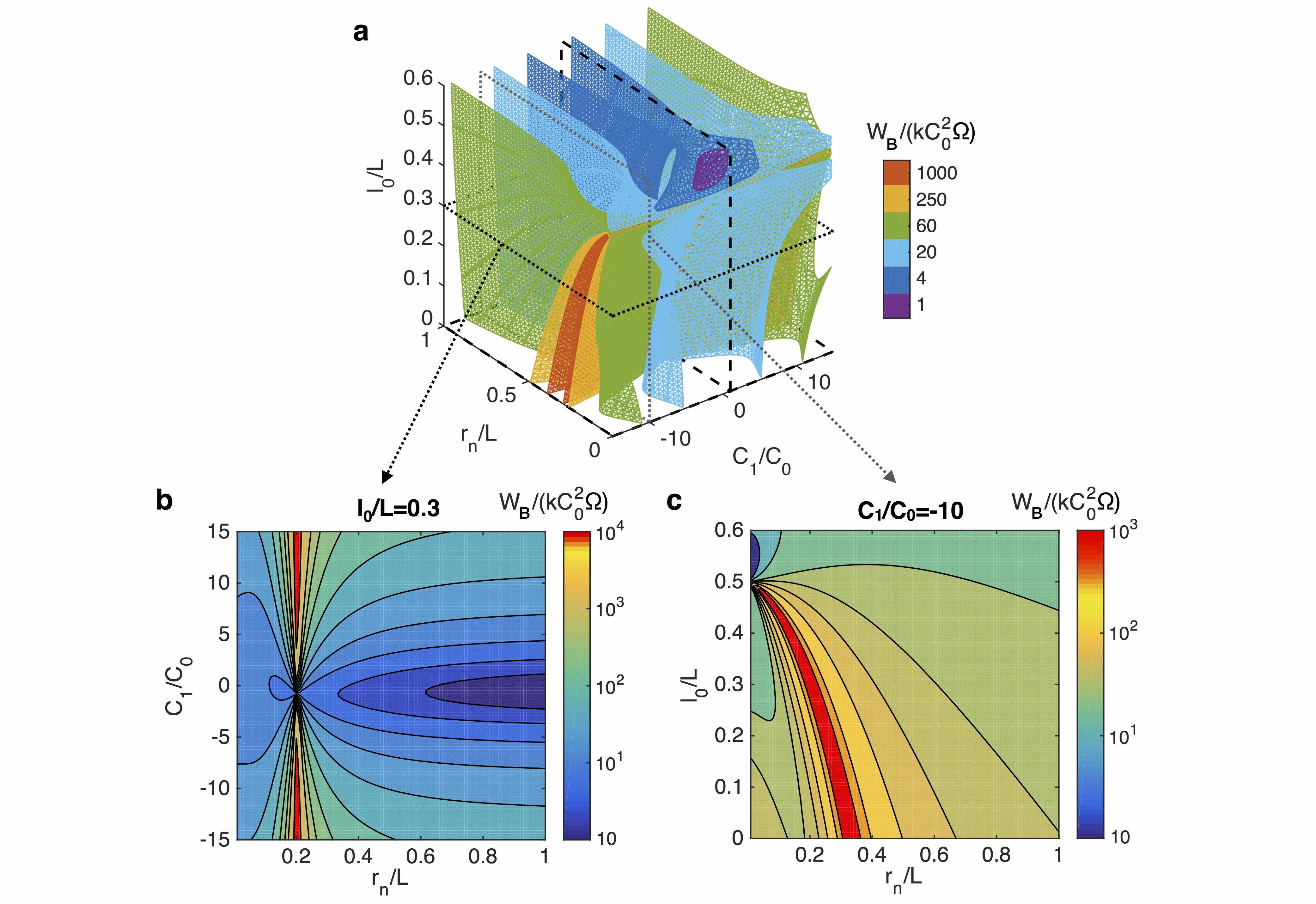}
\caption{Bending energy per unit area of catenoids in the parameters space defined by dimensionless neck radius ($r_n/L$), spontaneous curvature asymmetry ($C_1/C_0$), and geometrical asymmetry ($l_0/L$). (a) Isosurfaces of the energy. The two dashed planes correspond to the energy landscapes shown in Fig.~\ref{fig:cat_DC1}(e) and \ref{fig:cat_l}(e). The dotted planes correspond to energy landscapes for $l_0/L=0.3$ shown in (b), and $C_1/C_0=-10$ shown in (c).}
\label{fig:3d_phase}
\end{figure*}

Thus far, we have only considered catenoids that are geometrically symmetric, that is, both sides of the neck have equal arc length. Yet, asymmetric catenoids are more common in the cellular environment. For instance, the neck connecting a tube to a larger membrane reservoir is a catenoid partially truncated \cite{derenyi2002, jiang2008}. We therefore ask how does geometric asymmetry influence the distribution of spontaneous curvature and energy associated witht the switch?

We conduct simulations on truncated catenoids of arc length $L-l_0$ as shown in Fig.~\ref{fig:cat_l}(a), with both boundaries subject to the same spontaneous curvature $C_0$. The profile of spontaneous curvature along the arclength are shown for different degrees of geometrical asymmetry in Figs.~\ref{fig:cat_l}(b) and (c). We find that reducing asymmetry in the catenoid modifies the critical neck radius at which the switch in spontaneous curvature occurs. As seen from Figs.~\ref{fig:cat_l}(d) and (e), where the energy of the catenoid is plotted as a function of the neck radius, the larger the degree of asymmetry $l_0$, the smaller the critical neck radius at which the energy barrier occurs. 
Once half or more of the catenoid is cut off, corresponding to $l_0\geq L/2$, the energy barrier completely vanishes, allowing the neck radius to transition from large to small values through spontaneous curvature mechanisms only. 

For completeness, we fully explore all the combinations of spontaneous curvature differential and geometric asymmetry of the catenoid by computing the corresponding energy space. As shown in Fig.~\ref{fig:3d_phase}(a), the results confirm the behaviors described above, where the differential in $C$ across the neck mainly influences the width and intensity of the energy barrier, while the geometrical asymmetry determines the critical neck radius corresponding to the energy barrier and the switch. This is further shown in the two energy isovalue planes presented in Figs.~\ref{fig:3d_phase}(b) and (c) which have overall similar behaviors as those plotted in Figs.~\ref{fig:cat_DC1}(e) and \ref{fig:cat_l}(e) respectively.

\section*{Discussion}

Necks are ubiquitous in membrane biology, appearing as a necessary step in vesiculation processes and connecting tubules to membranes reservoirs. These structures can be studied and understood as catenoids \cite{kozlovsky2003, mcdargh2016}, which are minimal surfaces with zero mean curvature and negative Gaussian curvature everywhere. The formation of necks has been associated with line tension \cite{liu2006}, the change in Gaussian modulus \cite{baumgart2005} and other forces, but the interaction between spontaneous curvature and Gaussian curvature had not been explored until now. 

In this study, we explored the intricate relationship between Gaussian curvature, spontaneous curvature, and neck geometry. We asked, given a neck geometry connecting two reservoirs of curvature-inducing proteins, what spontaneous curvature field would satisfy the minimum energy requirement for a bilayer. We found a rather non-intuitive answer: the spontaneous curvature field depends on the Gaussian curvature and its intensity follows a switch-like behavior depending on the neck radius of the catenoid. The catenoid-shaped neck has an energy barrier at a critical neck radius corresponding to the switch in the sign of the  spontaneous curvature. We further identified two mechanisms allowing the modulation of this energy barrier -- (i) amplifying the spontaneous curvature differential at the boundaries increases the intensity and width of the energy barrier (Fig.~\ref{fig:cat_DC1}) and (ii) the geometrical asymmetry of the catenoid determines the critical neck radius at which the energy barrier is located (Fig.~\ref{fig:cat_l}). Moreover we found that the switching behavior is lost in specific cases: when the spontaneous curvature at the boundaries have opposite value (Fig.~\ref{fig:cat_DC2}), and when half or more of the catenoid is truncated (Fig.~\ref{fig:cat_l}).


Spontaneous curvature of lipid bilayers can be produced by a variety of relatively well understood mechanisms \cite{zimmerberg2006, jarsch2016, chabanon2017a}. In particular, the insertion of amphipathic $\alpha$-helix into lipid bilayers is known to induce curvature and is involved in several neck formation processes. The amphipathic $\alpha$-helixe is a conserved protein structure that can be found in Arf1 (involved in COP vesicle intracellular trafficking \cite{beck2011}), Epsin (involved in actin and clathrin mediated endocytosis \cite{boucrot2012}), and M2 proteins from influenza virus (involved in viral budding \cite{mishra2008, rossman2010, martyna2017}). All of these proteins have been shown to participate in membrane fission, which corresponds to the limiting shape of neck constriction \cite{beck2011, messa2014, martyna2017}. A similar mechanism is utilized by antimicrobial peptides (AMPs) to form buds and destabilize lipid membrane by inducing negative Gaussian curvature \cite{schmidt2010, schmidt2013, saikia2017}. 
While, to our knowledge, no explicit relationship have been proposed between such curvature-inducing proteins and spontaneous curvature, the knowledge of how spontaneous curvature is distributed on a catenoid-shaped neck is a first step in mapping the protein distribution that produce a membrane neck. A theoretical example of simple relationship between membrane protein and spontaneous curvature is given in the Supplementary Material.


One of the main findings of this study is the existence of an energy barrier for catenoid-shaped necks at a certain neck radius. This energy barrier is accompanied with a switch in the sign of spontaneous curvature along the catenoid. This behavior can be related to several biological mechanisms relevant to neck formation and membrane fission. Several studies have shown that lipids with negative spontaneous curvature are important during the fission process \cite{kooijman2005, churchward2008, schmidt2011, schmidt2013}. We found that a change in the sign of the spontaneous curvature is important to overcome the energy barrier associated with reducing neck size. Therefore, it is possible that by harnessing the heterogeneity of lipid species \cite{vanmeer2008}, and chemical reactions that can lead to the formation of lipids with negative spontaneous curvature, pre-fission structures overcome the energy barrier associated with necks. 

For instance, viral budding in influenza occurs in two main steps. First a neck is formed by a combination of scaffolding and lipid phase separation originating from the viral envelope. This step can only produce a neck of about 25 nm diameter, necessitating the action of another mechanism to further constrain the neck. During the second step, M2 amphipathic $\alpha$-helix insertion induces negative Gaussian curvature enabling the neck radius to reach values below 5 nm \cite{rossman2010, martyna2017}, at which point spontaneous membrane scission can occur \cite{campelo2012, liu2006}. It should be noted that no necking is possible with M2 amphipathic $\alpha$-helix only \cite{martyna2017}. The requirement for two distinct mechanisms in two regions of neck radii could be related to the energy barrier that we found as the neck radius of the catenoid decreases. Other evidence of switch-like behaviors are found in endocytosis where multiple studies have reported the existence of a snap-through instability \cite{walani2015, frolov2003, irajizad2017, hassinger2017}.  Furthermore, soap films, with no bending rigidity, also exhibit a structural instability when held as catenoids \cite{isenberg1978, powers2002}. Taken together, these findings suggest that perhaps, cellular membranes may utilize a fundamental geometric feature of catenoids to shape their membranes.

Our theoretical treatment results in two equations describing the equilibrium state of an elastic membrane conforming to a minimal surface: the shape equation (Eq.~\ref{eq:hape_mini}) that relates the membrane geometry (Gaussian curvature and surface Laplacian) to the spontaneous curvature, and the admissibility condition for the membrane tension (Eq.~\ref{eq:admiss_lambda}). These equations have been obtained by setting the mean curvature of the membrane to zero, which is a property of minimal surfaces.
\red{To further demonstrate the validity of our approach, we verified that the resulting geometry obtained by solving the general shape equation (without constraining the mean curvature) with the imposed field of spontaneous curvature from Fig.~\ref{fig:cat2}(c) are indeed catenoids. These results are presented in Section 3 of the Supplementary Material, and validate our approach.}

In the present model, we consider membrane deformations much larger than the bilayer thickness (see assumption 1), allowing us to treat the membrane as an infinitely thin surface. It is important however to note that for membrane structures barely larger than the membrane thickness, such as inverse cubic phases, the bilayer exhibits very large curvatures \cite{tran2015, demurtas2015}. In this case, phenomena such as thickness variation and tilt become important, and the mechanics of each individual monolayer must be taken into account \cite{siegel2004, wang2016, terzi2017}. It follows that the surface of interest with zero mean curvature is not the midplane of the bilayer anymore, but the neutral (or pivotal) plane, which is shifted from the midplane toward the water/lipid interface. Given that membrane necks studied here have characteristic sizes at least ten times larger than the bilayer thickness, we adopted the bilayer description where the midplane and neutral planes are identical. This assumption would not hold at stages close to membrane fission, requiring small scale modeling of the membrane. These issues have been discussed in detail in \cite{kozlovsky2002a, kozlovsky2003, kozlov2010}.

\red{
In general, the choice of the expression for the energetic contribution of the proteins to the membrane $A(\sigma)$ is a non-trivial endeavor. The energetic considerations for $A(\sigma)$ are discussed in \cite{steigmann2018} and specific examples provided in \cite{agrawal2011, callan-jones2016, belay2017}.  Models for entropic interactions of proteins with membranes have been explored in \cite{linden2012}, and the entropic contribution in multicomponent membranes have been studied in \cite{julicher1996, jiang2008, givli2012}. It is clear from these studies that more analyses between experiments and model equations are needed to identify the relevant constants and how $A(\sigma)$ may depend on the specific properties of the membrane proteins. In the case of minimal surfaces, the equations of motion decouple (Eqs.~\ref{eq:hape_mini} and \ref{eq:admiss_lambda}) allowing us to glean insight into $C(\sigma)$ without having to provide an explicit expression for $A(\sigma)$ and for $C(\sigma)$. 
}

\begin{figure*}[tbp]
\center
\includegraphics[width=\textwidth]{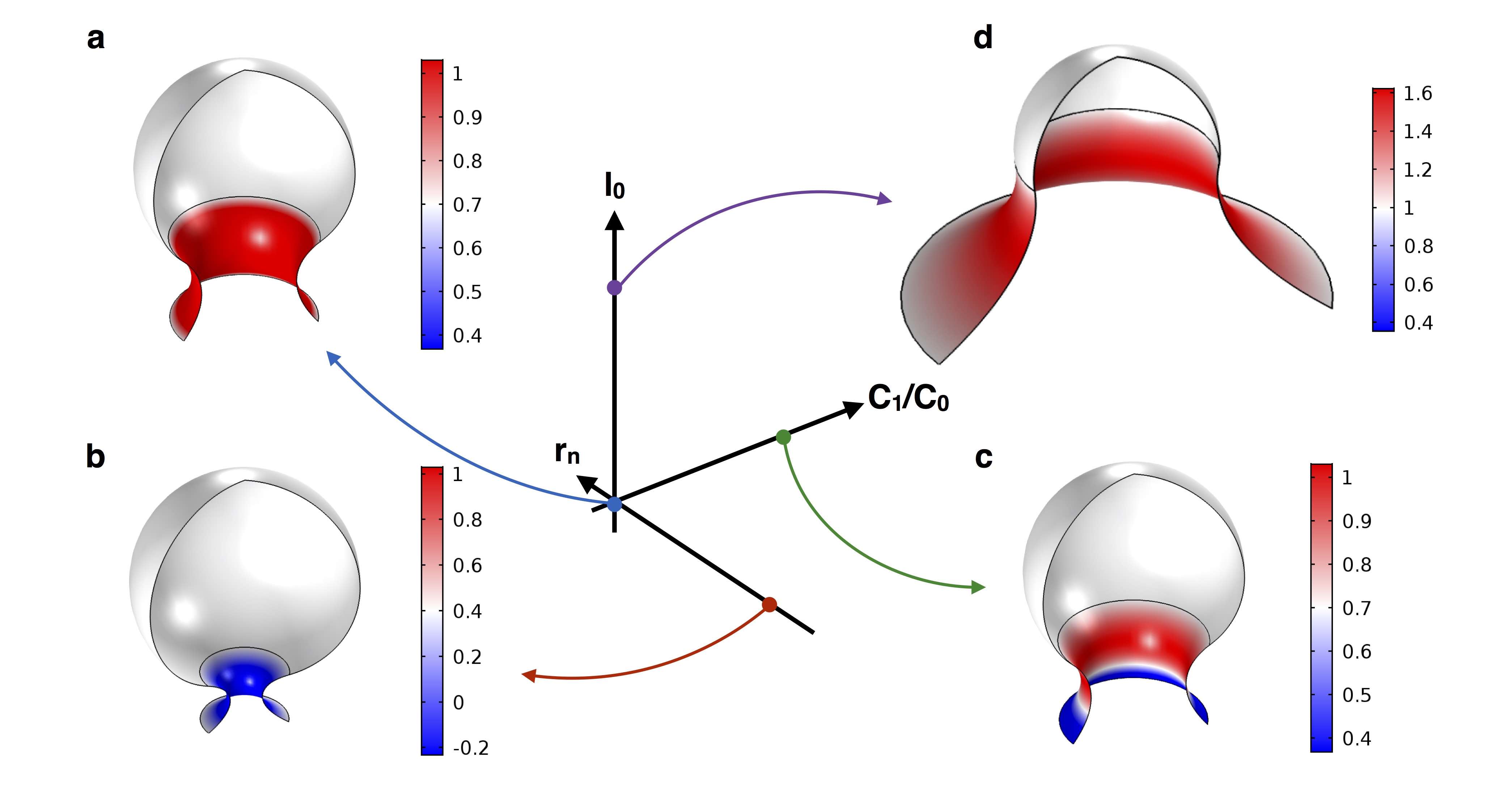}
\caption{Three features control the distribution of a spontaneous curvature along a catenoid-shaped neck, the neck radius, the boundary conditions, and the asymmetry of the catenoid. These can be utilized to form or maintain buds connected to a reservoir of curvature inducing proteins through a catenoid.}
\label{fig:buds}
\end{figure*}

While our model assumes an idealized catenoid shape for membrane necks, it is possible that necks may not remain as exact catenoids and the dynamics of the neck formation process, including biochemical reactions, heterogeneity in membrane composition and moduli \cite{hassinger2017}, forces exerted by proteins and cytoskeleton molecules, and in-plane diffusion of lipids and proteins \cite{agrawal2011, rangamani2013a, rangamani2014, bahmani2015} play an important role during fission and fusion. Despite these shortcomings, we have identified some fundamental features of the interaction between Gaussian curvature and spontaneous curvature in catenoids. We summarize our findings as a phase space where the spontaneous curvature, neck radius, and the geometric asymmetry of the catenoid can be altered to obtain buds and necks of different radii (Fig. \ref{fig:buds}). These variables might serve as design parameters for artificial membrane constructs and a stepping stone for further investigation of how membrane geometry and proteins interact.

\begin{small}
\bibliography{refs_minisurf}
\end{small}

\section*{Data availability}
The datasets generated and analyzed during the current study are available from the corresponding author on reasonable request.

\section*{Acknowledgments}
This work was supported in part by AFOSR FA9550-15-1-0124 and NSF PHY-1505017 awards to P.R. The authors thank Prof. David Steigmann for insightful discussions, and  Jasmine Nirody, Julian Hassinger, and Miriam Bell for critical reading of the manuscript.

\section*{Author Contributions}
Both authors developed the model, discussed the results, and wrote the article. M.C. performed the simulations.

\section*{Competing Financial Interests}
The authors declare no competing financial interests.


\pagebreak

\counterwithin*{equation}{section}
\renewcommand{\theequation}{S\thesection.\arabic{equation}} 
\renewcommand{\thefigure}{S\arabic{figure}} 

\section*{Supplementary material for: \\
\LARGE Gaussian curvature directs the distribution of spontaneous curvature on bilayer membrane necks}

\section*{Morgan Chabanon$^1$ and Padmini Rangamani $^{1,\star}$ }

\noindent $^1$ Department of Mechanical and Aerospace Engineering, University of California San Diego, La Jolla, CA, USA. \\
$^\star$ padmini.rangamani@eng.ucsd.edu

\bigskip

\section{Elastic lipid bilayer with one type of curvature-inducing protein}

In this section, we present a brief derivation of the generalized shape equation based on local stress balance. This approach, developped in details in \cite{steigmann1999, rangamani2013a, rangamani2014}, results in an equivalent model to the one obtained by the variationnal consideration \cite{steigmann2003, agrawal2009a, agrawal2009}. We then express specialize the model to minimal surfaces, and propose a relationship between spontaneous curvature and protein density. Finally we discuss the need for a multi-protein model.

\subsection{Equilibrium model of elastic surfaces by local force balance}

The equation of mechanical equilibrium of an elastic surface $\omega$ subject to a lateral pressure $p$ can be written in the compact form
\begin{equation} \label{eq:Sforce_balance}
\bm{\sigma}_{;\alpha}^\alpha + p \mathbf{n} = \mathbf{0} \; , 
\end{equation}
where $\bm{\sigma}^\alpha$ are the stress vectors and $\mathbf{n}$ is the unit normal to the local surface. Greek indices range over ${1,2}$, and if repeated, are summed over this range. Semicolon identifies covariant differentiation with respect to the metric $a_{\alpha\beta}=\mathbf{a}_\alpha \cdot \mathbf{a}_\beta$ where $ \mathbf{a}_\alpha=\mathbf{r}_{,\alpha}$ are the tangent vectors and $\mathbf{r}(\theta^\alpha)$ is the parametrization of the position field. The commas refer to partial derivatives with respect to the surface coordinates $\theta^\alpha$. With these definitions, the normal vector is given by $\mathbf{n} = (\mathbf{a}_1 \times \mathbf{a}_2) / \mid \mathbf{a}_1 \times \mathbf{a}_2 \mid$. In Eq.~\ref{eq:Sforce_balance}, the differential operation represents the surface divergence defined as
$
\bm{\sigma}_{;\alpha}^\alpha = (\sqrt{a})^{-1}(\sqrt{a} \bm{\sigma}^\alpha )_{,\alpha}
$
where $a=\det(a_{\alpha\beta})$. In surface theory, a manifold is described the metric $a_{\alpha\beta}$ defined above, and the curvature tensor given by $b_{\alpha\beta}=\mathbf{n}\cdot\mathbf{r}_{,\alpha\beta}$. 

For an elastic membrane whose energy surface density per unit mass depends on the metric and curvature only $F(a_{\alpha\beta}, b_{\alpha\beta}; \theta^\alpha)$, the stress vectors involved in the local force balance (Eq.~\ref{eq:Sforce_balance}) can be written as \cite{steigmann1999}
\begin{equation} \label{eq:SAsig}
\bm{\sigma}^\alpha = \mathbf{T}^\alpha + S^\alpha\mathbf{n} \;.
\end{equation}
Here the tangential stress vectors are
\begin{equation}  \label{eq:SAT}
\mathbf{T}^\alpha = T^{\beta\alpha} \mathbf{a}_\beta \quad \text{with} \quad
T^{\beta\alpha} = \sigma^{\beta\alpha} + b_\mu^\beta M^{\mu\alpha} \;,
\end{equation}
and the components of the normal stress vectors are
\begin{equation} \label{eq:SAS}
S^\alpha = -M_{;\beta}^{\alpha\beta} \;,
\end{equation}
where $b_\alpha^\beta = a^{\beta\lambda} b_{\lambda\alpha}$. The components of the stress vectors depends on the energy density as \cite{steigmann1999}
\begin{equation} \label{eq:SsigmaM}
\sigma^{\alpha\beta} = \rho \left( \frac{\partial F}{\partial a_{\alpha\beta}} + \frac{\partial F}{\partial a_{\beta\alpha}} \right) \quad \text{and} \quad  
M^{\alpha\beta} = \frac{\rho}{2} \left( \frac{\partial F}{\partial b_{\alpha\beta}} + \frac{\partial F}{\partial b_{\beta\alpha}} \right) \;,
\end{equation}
where $\rho$ the surface mass density of the membrane. The tangential and normal local force balances can now be obtained by introducing Eqs.~\ref{eq:SAsig}, \ref{eq:SAT}, and \ref{eq:SAS} into Eq.~\ref{eq:Sforce_balance}, resulting in
\begin{equation} \label{eq:Stan_norm_force_bal}
T_{;\alpha}^{\beta\alpha} - S^\alpha b_\alpha^\beta = 0  \quad \text{and} \quad
S_{;\alpha}^\alpha + T^{\beta\alpha}b_{\beta\alpha} +p =0 \;,
\end{equation}
where we made use of the Gauss and Weingarten equations \cite{sokolnikoff1964} $\mathbf{a}_{\alpha;\beta} = b_{\alpha\beta} \mathbf{n}$ and $\mathbf{n}_{,\alpha}=-b_\alpha^\beta\mathbf{a}_\beta$ respectively.

Practically, the free energy density is sometimes given as a function of the mean curvature $H$ and Gaussian curvature $K$. These are related to the metric and curvature by
\begin{equation} \label{eq:SHK_fundamental}
H = \frac{1}{2} a^{\alpha\beta} b_{\alpha\beta} \; , \quad 
K = \frac{1}{2} \varepsilon^{\alpha\beta}\varepsilon^{\lambda\mu} b_{\alpha\lambda}  b_{\beta \mu} \; ,
\end{equation}
where $a^{\alpha\beta} = (a_{\alpha\beta})^{-1} $ is the dual metric, and $\varepsilon^{\alpha\beta}$ is the permutation tensor defined by $\varepsilon^{12}=-\varepsilon^{21}=1/\sqrt{a}$, $\varepsilon^{11} = \varepsilon^{22} =0$. According to the definitions \ref{eq:SHK_fundamental}, the free energy density per unit mass can be re-written in terms of the mean and Gaussian curvature $F(H,K; \theta^\alpha)$. Furthermore, lipid membranes are essentially incompressible (see assumption 3 above). This leads us to introduce a Lagrange multiplier $ \gamma(\theta^\alpha)$ to ensure that the local area dilatation $J=1$, or equivalently, to constraint the constant surface density $\rho$ of the membrane. Consequently we can define the surface energy density of the membrane as follows
\begin{equation}
F(\rho, H, K; \theta^\alpha) =  \bar{F}(H,K; \theta^\alpha) - \frac{\gamma(\theta^\alpha)}{\rho} \;,
\end{equation}
and when introducing the surface energy per unit area $W(\rho, H,K; \theta^\alpha) = \rho \bar{F}(H,K; \theta^\alpha)$, the components of the stress vectors (Eqs.~\ref{eq:SsigmaM}) can be written as \cite{steigmann1999}
\begin{gather} \label{eq:SsigHK}
\sigma^{\alpha\beta} = (\lambda+W) a^{\alpha\beta} - \left( 2HW_H + 2KW_K \right) a^{\alpha\beta} + W_H \tilde{b}^{\alpha\beta} \\ \label{eq:SMHK}
M^{\alpha\beta} = \frac{1}{2}W_H a^{\alpha\beta}  +  W_K \tilde{b}^{\alpha\beta}
\end{gather}
where $\lambda(\theta^\alpha) = - \left[ \gamma(\theta^\alpha) + W(H, K; \theta^\alpha) \right]$, and $\tilde{b}_{;\beta}^{\alpha\beta} = 2H a^{\alpha\beta} - b^{\alpha\beta}$ is the cofactor of the curvature. The subscripts $H$ and $K$ refer to the partial derivative of the energy with respect to the indicated variable. The Lagrange multiplier $\gamma$ has a mechanical interpretation of surface pressure and is not a material property of the surface \cite{rangamani2013a, rangamani2014}. $\lambda$ can be interpreted as the surface tension based on comparisons with edge conditions on a flat surface \cite{rangamani2014}.

Finally, introducing Eqs.~\ref{eq:SsigHK} and \ref{eq:SMHK} into Eqs.~\ref{eq:SAT} and \ref{eq:SAS}, we can rewrite the normal and tangential force balances (Eqs.~\ref{eq:Stan_norm_force_bal}) as
\begin{equation} \label{eq:Sgen_shape}
\Delta \left( \frac{1}{2}W_H \right) + (W_K)_{;\alpha\beta} \tilde{b}^{\alpha\beta} + W_H (2H^2 -K) + 2H(KW_K - W) = p+2\lambda H \;,
\end{equation}
and 
\begin{equation} \label{eq:Sgen_incomp}
-\left( \gamma_{,\alpha} + W_K K_{,\alpha} + W_H H_{,\alpha} \right) a^{\beta\alpha} =
\left( \frac{\partial W}{\partial \theta^\alpha}\mid_\text{exp} + \lambda_{,\alpha} \right) a^{\beta\alpha} =0 \;,
\end{equation}
where $\Delta(\cdot) = (\cdot)_{;\alpha\beta} a^{\alpha\beta}$ is the surface Laplacian (or Beltrami operator), and  $\partial(\cdot) / \partial \theta^\alpha\mid_\text{exp}$ is the explicit derivative with respect to $\theta^\alpha$.

Eqs.~\ref{eq:Sgen_shape} and \ref{eq:Sgen_incomp} are the general shape equation and incompressibility condition for an elastic surface with free energy per unit area $W(\rho, H, K; \theta^\alpha)$. In the following we specialize it to the case of lipid membranes by specifying the form of the free energy.

\subsection{Elastic lipid bilayers with non-constant spontaneous curvature}

The most common model of lipid membranes is the Helfrich energy \cite{helfrich1973}. This can be extended to account for the entropic contribution of membrane-bound proteins to the areal free-energy functional such as
\begin{equation} \label{eq:SWdiff} 
W(\sigma, H,K; \theta^\alpha)=A(\sigma )+k(\theta^\alpha)[H-C(\sigma )]^{2} + k_G(\theta^\alpha) K \;,
\end{equation}%
Here $A(\sigma)$ is the contribution of the membrane-bound proteins to the free energy and $\sigma$ is the surface density of proteins. $k(\theta^\alpha)$ and $k_G(\theta^\alpha)$  are the bending and Gaussian moduli respectively, considered to be surface coordinate dependent. $C(\sigma )$ is the spontaneous (mean) curvature, which is determined by the local membrane composition, say the surface density of a curvature-inducing protein $\sigma$. We will propose later a possible relationship for $C(\sigma)$.
While it is certainly possible to propose explicit functions of $A(\sigma)$ and $C(\sigma)$ on the protein density (see \cite{steigmann2018} for discussion on $A(\sigma)$, and \cite{agrawal2011, belay2017} for specific examples) we will for now retain their general form.

The shape equation for lipid membrane with protein and space dependent moduli is obtained by introducing the free energy density \eqref{eq:SWdiff} into Eq.~\eqref{eq:Sgen_shape}, resulting in 
\begin{equation} \label{eq:Sshape}
\Delta \left[ k (H-C) \right] + 2H\Delta k_G - (k_G)_{;\alpha\beta} b^{\alpha\beta} + 2k(H-C)(2H^2-K) + 2H(k_G K - W(\sigma,H,K;\theta^\alpha) ) = p + 2\lambda H \;.
\end{equation}
The incompressibility for lipid membranes is obtained similarly, introducing \eqref{eq:SWdiff} into the Eq.~\eqref{eq:Sgen_incomp}
\begin{equation}\label{eq:Sincomp}
\nabla \lambda = -W_\sigma \nabla \sigma
-\nabla k (H-C)^2 - \nabla k_G K \;,
\end{equation} 
where $(\cdot)_\sigma = \partial (\cdot) / \partial \sigma$ is the partial derivative with respect to $\sigma$, and $\nabla (\cdot) = (\cdot)_{,\alpha} a^{\alpha\beta}$ is the surface gradient. One can recognize $W_\sigma$ as the chemical potential of the membrane protein, and given Eq.~\eqref{eq:SWdiff}, we have
\begin{equation} \label{eq:SWsigma}
W_\sigma = A_\sigma - 2k(H-C)C_\sigma \;.
\end{equation} 

Eqs.~\ref{eq:Sshape} and \ref{eq:Sincomp} describe the equilibrium configuration of lipid membrane subject to heterogeneous spontaneous curvature induced by proteins. An additional constraint related to the area incompressibility of the membrane requires the lipid velocity field ($\mathbf{u} = u^\alpha \mathbf{a}_\alpha + w \mathbf{n}$) to satisfy \cite{agrawal2011}%
\begin{equation} \label{eq:Sincomp_u}
u_{;\alpha }^{\alpha }=2Hw \;.
\end{equation}%

Although models for lipid flow within biological membranes have been proposed \cite{rangamani2013a, rangamani2014, bahmani2015, steigmann2018}, such description is out of the scope of this study. Provided a lipid velocity field satisfying Eq.~\ref{eq:Sincomp_u}, and suitable boundary conditions, the system given by the coupled equations \ref{eq:Sshape} and \ref{eq:Sincomp} fully describes the equilibrium configuration of a lipid membrane subject to a static distribution of curvature-inducing proteins. 

\subsection{Static distribution of curvature-inducing protein on minimal surfaces}

In this section, we specialize the system of Eqs.~\ref{eq:Sshape} and \ref{eq:Sincomp} to minimal surfaces, and examine the associated restrictions on the Lagrange multiplier field $\lambda$ and velocity field $u^\alpha$.

Before to proceed, it is useful to clarify what are the imposed and the unknown quantities in our model. Traditionally, one seeks to compute the shape of the membrane for a given distribution of spontaneous curvature and boundary conditions. This is often done by formulating the shape equation and incompressibility condition within a certain parametrization. For instance, within the Monge parametrization one would aim to compute the height of the membrane $h(x,y)$ with respect to a reference plane at every point, while in axisymetric coordinates one would solve for the distance to the axis of symmetry $r(z)$. In our case however, we consider the \textit{inverse problem}, that is to say, we seek to compute the distribution of spontaneous curvature for a given shape of the membrane and boundary conditions. Independently of the approach, the spontaneous curvature $C(\sigma)$ is interpreted physically as resulting form the distribution of curvature-inducing proteins (or lipids) of areal density $\sigma$ on the membrane. We illustrate our `inverse problem' approach by choosing minimal surfaces as the imposed membrane shape. We further demonstrate the applicability of our model by solving on catenoid-like necks, which are minimal surfaces that have been extensively used as models for the study of fusion/fission intermediate.

Minimal surfaces are characterized by the property that the mean curvature
vanishes pointwise ($H=0$ everywhere on the membrane). Furthermore, as a first approximation, we consider membranes with isotropic mechanical properties ($k$ and $k_G$ are constants). Accordingly, in the absence of transmembrane pressure, the shape equation \ref{eq:Sshape} reduces to a variable-coefficient Helmhotz equation for the spontaneous curvature
\begin{equation} \label{eq:Sshape_mini}
\Delta C(\sigma)-2KC(\sigma)=0 \;.
\end{equation}%
Both the energetic contribution of the proteins $A(\sigma)$ and the local Lagrange multiplier $\lambda$ are now absent from the shape equation \ref{eq:Sshape_mini}, therefore uncoupling them from the incompressibility condition. Yet, any solution of Eq.~\ref{eq:Sshape_mini} is restricted to the condition that the balance equation \ref{eq:Sincomp} with \ref{eq:SWsigma} is satisfied. For minimal surfaces, these later equations reduce to
\begin{equation}
\nabla \lambda = -\left[A(\sigma )_\sigma+2kC(\sigma )C(\sigma )_\sigma \right] \nabla \sigma \;.
\end{equation} 
Using the identities $\nabla A(\sigma )=A(\sigma )_\sigma\nabla \sigma$ and  $\nabla \left[C(\sigma )^2\right]=2 C(\sigma) C(\sigma )_\sigma \nabla \sigma$, this can be simplified to
\begin{equation}
\nabla \lambda =-\nabla \lbrack A(\sigma )+kC(\sigma )^{2}] \;,
\end{equation}%
from which we get $\lambda $ as a function of $A(\sigma)$ and $C(\sigma)$ apart from a
constant $\lambda_0$, such that
\begin{equation} \label{eq:Sadmiss_lambda}
\lambda =-[A(\sigma )+kC(\sigma )^{2}] + \lambda_0 \;.
\end{equation}
Equation \ref{eq:Sadmiss_lambda} is the admissibility condition for the Lagrange multiplier field $\lambda$.

\subsection{Relationship between protein density and spontaneous curvature} \label{sec:Csigma}

Here we propose an explicit relationship between the spontaneous curvature and the distribution of curvature-inducing proteins. 

\paragraph{One protein model.}

Let us consider a pointwise protein surface density on the membrane $\sigma(\theta^\alpha)$. The goal of this section is to propose an expression for $C(\sigma)$. Note that we consider proteins only for ease of visualization, but our model can be as easily applied to spontaneous curvature-inducing lipids or nano-objects instead of proteins.

\begin{figure*}[t!b]
\center
\includegraphics[width=\textwidth]{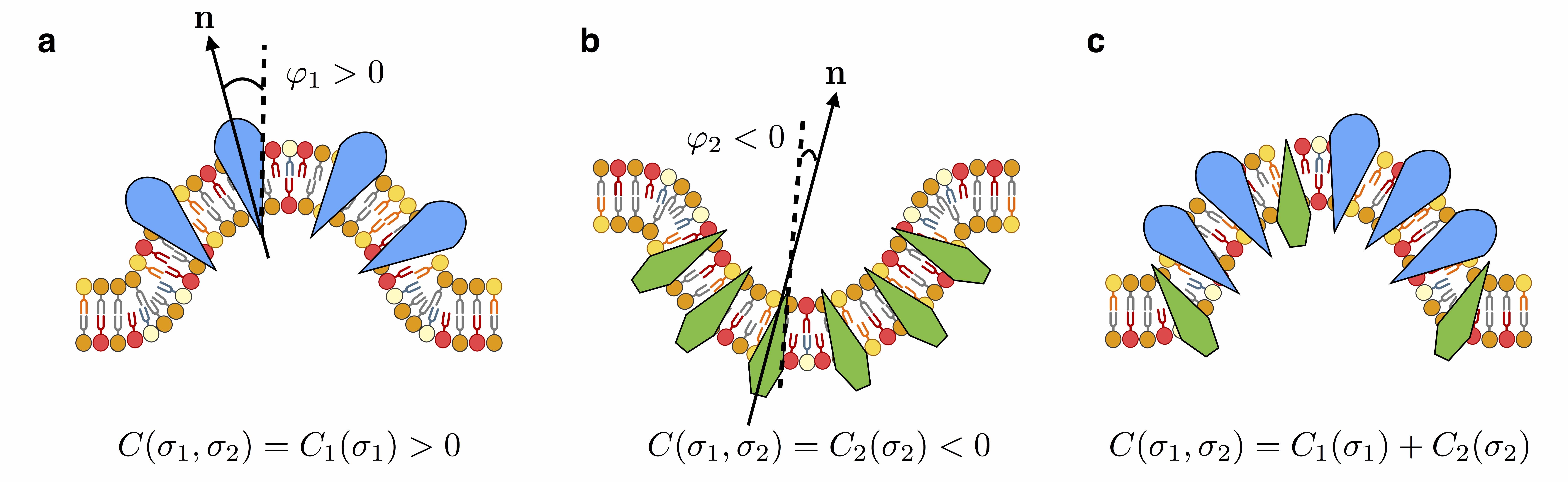}
\caption{Schematic representation of curvature-inducing proteins with conical shape on a lipid membrane. (a) Proteins with positive angle $\varphi$ induce positive spontaneous curvature. (b) Proteins with a negative angle $\varphi$ induce negative spontaneous curvature. (c) The value of the spontaneous curvature is the a combination of the local protein composition.
}
\label{fig:angle}
\end{figure*}

A convenient way to think about the relationship between spontaneous curvature and protein density, is in terms of the insertion of a conical shape transmembrane protein with its axis of revolution directed along the surface normal (see Fig.~\ref{fig:angle}). Following this representation, the point value of $C(\sigma)$ will depend on (i) the angle of the cone ($\varphi$), (ii) the lipid-protein specific moietic interactions ($\kappa$), and (iii) the local density of protein ($\sigma$). It should be noted that in this model, we neglect any thickness variation or lipid tilt resulting from the insertion of the proteins in the lipid bilayer. Although experimental observations seem to support point (iii) \cite{baumgart2011,zhu2012}, to our knowledge, no explicit relation between $C$ and $\sigma$ has been reported based on experimental data. Consequently, we consider a simple expression for the spontaneous curvature of the form \cite{agrawal2011}
\begin{equation} \label{eq:SCsigma1}
C(\sigma )=\kappa \varphi\sigma\;.
\end{equation}%
This form ensures that the induced spontaneous curvature vanishes in the case of cylindrical embedded proteins ($\varphi = 0$), and assumes that a same type of conical protein inserted from one or the other lipid leaflet will have $\varphi$ of opposite sign, and therefore produce a spontaneous curvature of opposite sign. It should be noted that this expression does not account for protein-protein interaction, and therefore should be considered only in a dilute regime, outside of any saturation effect.

A limitation of the model in its current form is that the spontaneous curvature is the result of only one type of protein. As a consequence, $C(\sigma)$ can only be either positive everywhere, or negative everywhere, but cannot change sign from one point of the membrane to the other. This is in contradiction with our results on the determination of spontaneous curvature on catenoid-shaped necks. The minimal set of proteins to allow the spontaneous curvature to have positive and negative values is two, and this case is considered next.

\paragraph{Two proteins model.}

We propose that the local protein density results from a minimal set of two proteins $\sigma = \{ \sigma_1, \sigma_2 \}$. All the results below can be easily generalized to a set of $N$ proteins. As a first approximation, we assume that the spontaneous curvature is a linear combination of the spontaneous curvatures induced by each individual proteins $C(\sigma_1,\sigma_2) = C_1(\sigma_1) + C_2(\sigma_2) $.
Similarly to the one protein model (Eq.~\ref{eq:SCsigma1}), we assume a linear relationship between spontaneous curvature and protein density, such that
\begin{equation}
C_i(\sigma_i )=(\kappa_i \varphi_i )\sigma_i \;,
\end{equation}%
where $\kappa_i $ is a positive constant representing the lipid/protein specific hydorphobic interactions, and $\varphi_i$ is the angle made by the meridian of the conic protein with the surface normal (see Fig.~\ref{fig:angle}). Note that in order for the total spontaneous curvature $C(\sigma_1,\sigma_2)$ to have either positive or negative values, the angles of the two proteins needs to have opposite signs ($\varphi_1 \varphi_2 < 0$).

With these forms in effect, Eqs.~\ref{eq:Sshape_mini} and \ref{eq:Sadmiss_lambda} can be expressed directly as a function of the protein densities. The shape equation thus reduces to
\begin{equation} \label{eq:Sshape_mini_s}
\Delta (\kappa_1 \varphi_1\sigma_1 + \kappa_2 \varphi_2\sigma_2) -2K(\kappa_1 \varphi_1\sigma_1 + \kappa_2 \varphi_2\sigma_2) =0 \;,
\end{equation}%
subject to the admissibility condition for $\lambda$
\begin{equation} \label{eq:Sadmiss_lambda_mini_s}
\lambda = -[ A(\sigma_1, \sigma_2) + k(\kappa_1 \varphi_1\sigma_1 + \kappa_2 \varphi_2\sigma_2)^2] + \lambda_0 \;.
\end{equation}

In the case where the two proteins have the same physical properties, but are inserted on either side of the membrane ($ \kappa_1= \kappa_2 = \kappa$, and $\varphi_1= - \varphi_2 = \varphi$, we can define the effective protein density $\tilde{\sigma} = \sigma_1 -\sigma_2$ which satisfies
\begin{equation}
\Delta \tilde{\sigma} -2K\tilde{\sigma} =0 \;,
\end{equation}%
and
\begin{equation} 
\lambda = -[A(\sigma_1, \sigma_2) + k (\kappa \varphi \tilde{\sigma})^2] + \lambda_0 \;.
\end{equation}

\section{Simple Oscillator Analogy}

\begin{figure*}[t!b]
\center
\includegraphics[width=\textwidth]{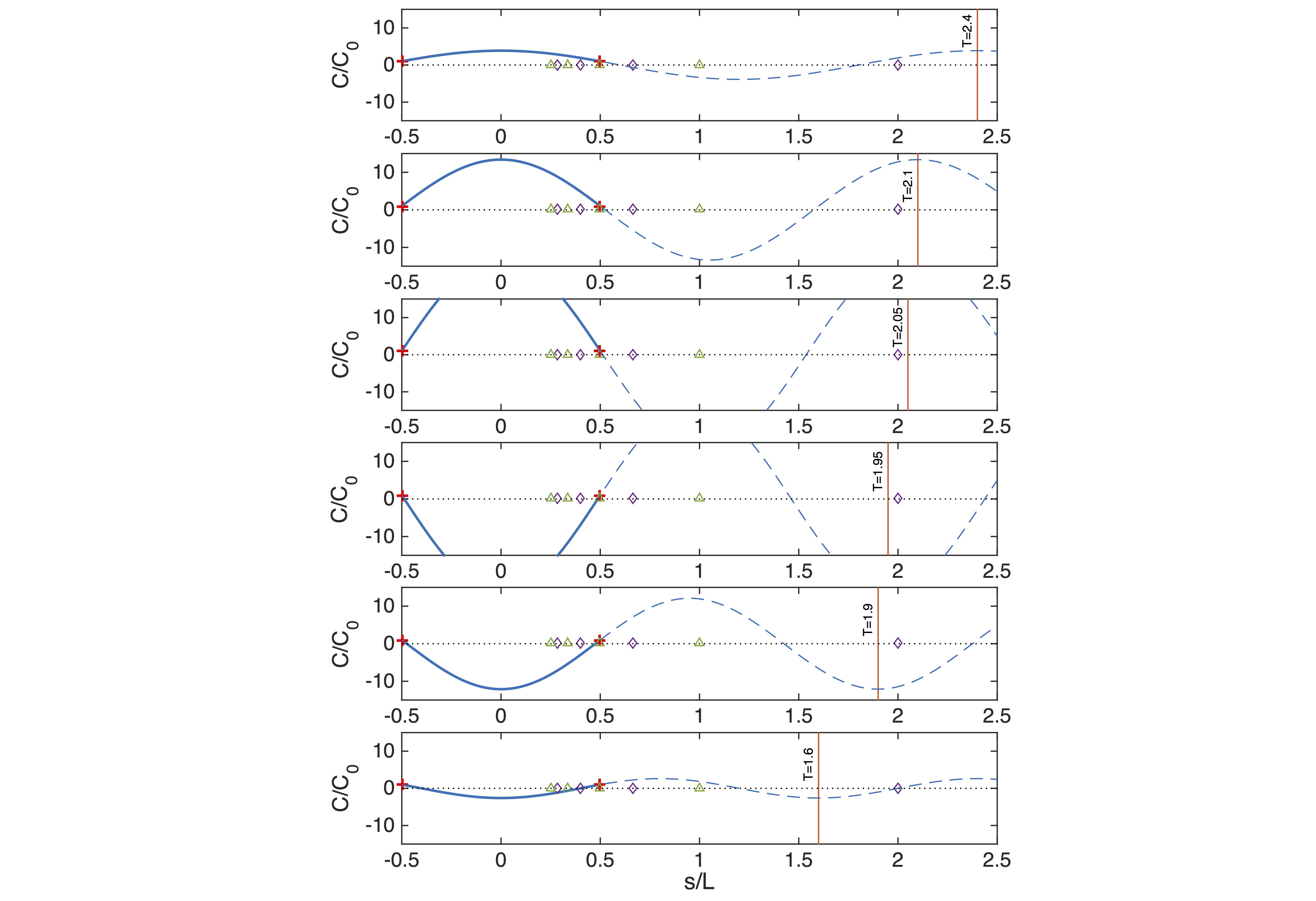}
\caption{The switch-like behavior in spontaneous curvature observed for catenoids can be conceptually understood with a simple oscillator. The solution of a simple harmonic oscillator (Eq.~\ref{eq:SC_oscillator}) with $C_1=C_0$ depends on the oscillator period $T= -\pi/K$. Decreasing values of the oscillator period correspond to increasing values of $K$, and therefore decreasing values of the neck radius. $C$ diverges for $T^*/L = 2/(1+2n)$ (marked by purple diamonds), and changes mode for $T^* /L= 1/(1+n)$ (indicated by green triangles).
}
\label{fig:oscillator}
\end{figure*}

In order to study the behavior of Eq.~\ref{eq:Sshape_mini} in simplified conditions, let us consider the case where $K$ is a constant. In one-dimension, Eq.~\ref{eq:Sshape_mini}  can be written as
\begin{equation}
\frac{d^2C}{ds^2} = - \omega^2 C \;,
\end{equation}
where $\omega^2=-2K$ is a positive constant, and $s\in[-L/2; L/2]$. This is the equation of a simple harmonic oscillator of period $T=2\pi/\omega = \pi \sqrt{-2/K}$, which has for  general solution
\begin{equation}
C(s) = A \cos(\omega s) + B \sin(\omega s) \; ,
\end{equation}
$A$ and $B$ being constants determined by the boundary conditions.
With boundary conditions $C(-L/2) = C_0$  and $C(L/2) = C_1$ the solution is
\begin{equation} \label{eq:SC_oscillator}
C(s) = \frac{C_0+C_1}{2\cos(\omega L /2)} \cos(\omega s) -  \frac{C_0-C_1}{2\sin(\omega L /2)} \sin(\omega s) \;.
\end{equation}
From Eq.~\ref{eq:SC_oscillator} we have that the value of $C$ at the neck ($s=0$) is
\begin{equation} \label{eq:SC0_oscillator}
C(0) = \frac{C_0+C_1}{2\cos(\omega L/2)} \;,
\end{equation}
which diverges for $\omega^*  = (\pi + 2n\pi)/L$, where $n\in \mathbb{N}$. Or in terms of the oscillator period, the solution diverges for
\begin{equation}
T^* = \frac{2L}{1+2n} \;.
\end{equation}
Eq.~\ref{eq:SC_oscillator} is plotted in Fig.~\ref{fig:oscillator} for various periods $T$. The value of $C$ within the interval $[-L/2;L/2]$ is positive for $T>T^*(n=1)$, and negative below. To decrease the oscillator period is conceptually equivalent to increase the absolute value of $K$, or to decrease the neck radius of the catenoid.  For a catenoid, the Gaussian curvature at the neck is $K(s=0)=-1/r_n^2$. Taking $\omega^2 = 2/r_n^2$, the positive value of the neck radius for which the spontaneous curvature diverges is
\begin{equation} \label{eq:Srn_oscillator}
r_n^* =  \frac{ \sqrt{2}L} {\pi(1+2n)} \;.
\end{equation}
For $n=0$, we have $r_n^*/L\simeq 0.45$. From Eq.~\ref{eq:Srn_oscillator}, it is clear that the value of the critical neck radius is independent of the boundary conditions.

\section{Direct solution of the shape equation}

In this section we verify that the solution for the spontaneous curvature obtained on catenoid-shaped neck does produce catenoid by following the ``direct'' approach. The shape equation for an isotropic membrane is
\begin{equation} \label{eq:Sshape_iso}
\Delta \left[ k (H-C) \right]  + 2k(H-C)(H^2-K+HC ) = p + 2H(\lambda +A)\;,
\end{equation}
with the incompressibility condition that can be written as
\begin{equation} \label{eq:Sincomp_iso}
\nabla( \lambda +A)=   2k(H-C)C_\sigma  \nabla \sigma \;.
\end{equation}

\subsection{Axisymmetric parametrization}

We write the equilibrium equations of the membrane in axisymmetric coordinates. We therefore define a surface of revolution that is described in the coordinate basis $(\mathbf{e}_r,\mathbf{e}_\theta, \mathbf{k})$ by
\begin{equation}
\mathbf{r} = r(s) \mathbf{e}_r + z(s) \mathbf{k} \;,
\end{equation}
where $s$ is the arclength along the curve, $r(s)$ is the radius to the axis of revolution, and $z(s)$ is the elevation from a reference plane. Since $r(s)^2 + z(s)^2 =1$, it is convenient to define the angle $\psi$ such that
\begin{equation}
\mathbf{a}_s = \cos \psi \mathbf{e}_r + \sin \psi \mathbf{k} \quad \text{and} \quad
\mathbf{n}= -\sin \psi \mathbf{e}_r + \cos \psi \mathbf{k} \;,
\end{equation}
are the tangent and normal vectors to the curve respectively. It follows that the surface can be parametrized as
\begin{gather} \label{eq:Sr_p}
r'(s) = \cos \psi \;, \\ \label{eq:Sz_p}
z'(s) = \sin \psi \;,
\end{gather}
where $(\cdot)' = d(\cdot)/ds$. We can now write the tangential and transverse principal curvatures as
\begin{equation}
\kappa_1 = \psi' \quad \text{and} \quad \kappa_2 = r^{-1} \sin \psi \;,
\end{equation}
respectively, and the mean and Gaussian curvatures as
\begin{gather}  \label{eq:Saxi_H}
H =\frac{\kappa_1 + \kappa_2}{2} = \frac{\psi' + r^{-1} \sin \psi}{2}    \\ \label{eq:Saxi_K}
 K=\kappa_1 \kappa_2 = H^2 -(H-r^{-1} \sin \psi )^2 \;,
\end{gather}
respectively. Eq.~\ref{eq:Saxi_H} provides the differential equation for $\psi$, which can be rearranged as
\begin{equation} \label{eq:Spsi_p}
r \psi' = 2 r H - \sin \psi \;.
\end{equation}

Eq.~\ref{eq:Sshape_iso} is a second order partial  differential equation. In order to simplify its resolution, we define $\Lambda$ as
\begin{equation} \label{eq:SLambda_iso}
\Lambda =  r\left[ k(H-C)\right]' \;,
\end{equation}
allowing to write the shape equation (Eq.~\ref{eq:Sshape_iso}) as a first order differential equation for the mean curvature
\begin{equation} \label{eq:SH_p}
H' = r^{-1}\Lambda + C'  \;.
\end{equation}

Using the relation $\nabla (H-C) = r^{-1} [(H-C)']' = r^{-1} (r L)'$, into Eq.~\ref{eq:Sshape_iso}, we get a differential equation for $\Lambda$
\begin{equation} \label{eq:SLambda_p}
r^{-1} \Lambda' = \frac{p}{k} + 2H\left[ (H-C)^2 + \frac{\lambda+A}{k} \right] - 2[H-C] \left[ H^2 + (H-r^{-1} \sin \psi )^2 \right] \;.
\end{equation}

Finally, Eq.~\ref{eq:Sincomp_iso} becomes
\begin{equation} \label{eq:Slambda_p}
[\lambda + A]' = 2k(H-C)C' \;.
\end{equation}

The full system is composed of Eqs.~\ref{eq:Sr_p}, \ref{eq:Sz_p}, \ref{eq:Spsi_p}, \ref{eq:SH_p}, \ref{eq:SLambda_p}, and \ref{eq:Slambda_p}, and must be completed with 6 boundary conditions. For a domain defined as $s \in [0 , L/2]$, we set
\begin{equation}
\begin{array}{l}
r(0) = r_n \;,\\
z(0) = 0 \;,\\
\psi(0) = \pi/2 \;,
\end{array} \quad
\begin{array}{l}
r(L/2) = r_n \cosh [z(L/2)/r_n] \;, \\
z(L/2) = r_n \text{asinh} [L/2/r_n] \;, \\
\psi(L/2) = \text{asin} [ 1/\sqrt{1+(L/2/r_n)^2} ] \;,
\end{array}
\end{equation}
where we used Eq.~\ref{eq:Saxi_K} to determine $\psi(L/2)$.

\subsection{Non-dimensional system}

\begin{figure*}[tb]
\center
\includegraphics[width=\textwidth]{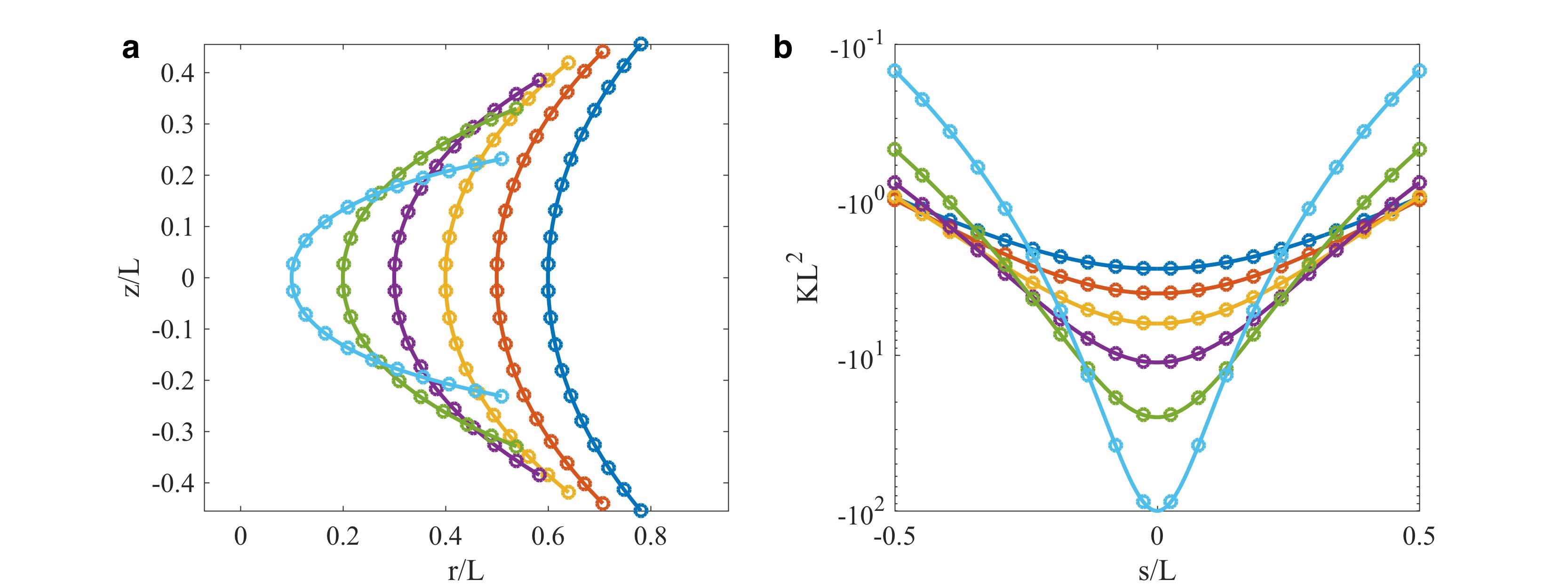}
\caption{Direct solution of the shape equation subject to the distribution of spontaneous curvature from Fig.~3(c). Symbols are exact values for a catenoid, solid lines are computation results. (a) Shape of the neck, and (b) Gaussian curvature.
}
\label{fig:direct}
\end{figure*}

Based on the length scales introduced in the main text, $L$ and $C_0$, we define the axisymmetric dimensionless variables as
\begin{equation}
\begin{array}{c}
\bar{s} = s/L \;, \quad 
\bar{r} = r/L \;, \quad 
\bar{z} = z/L \;, \quad 
\bar{H} = H L \;, \quad 
\bar{C} = C/C_0\;, \quad 
\bar{l} =  \Lambda L \;, \\
\bar{\lambda} = (\lambda + A) L^2/k \;, \quad 
\bar{p} = p L^3/k  \;, \quad 
\bar{K} = K L^2\;, \quad 
\bar{k}_G = k_G/k \;.
\end{array}
\end{equation}
The dimensionless system then becomes
\begin{equation} \label{eq:Snd_pde_axi}
\begin{array}{c}
\bar{r}' = \cos \psi \;, \quad
\bar{z}' = \sin \psi \;, \quad
\bar{r} \psi' = 2 \bar{r} \bar{H} - \sin \psi \;, \quad
\bar{H}' = \bar{r}^{-1} \bar{l} + \bar{C}' \;, \quad
\bar{\lambda}' = 2 (\bar{H} - \bar{C}) \bar{C}' \;,\\
\bar{r}^{-1} \bar{l}' = \bar{p} + 2 \bar{H} \left[ (\bar{H} - \bar{C})^2 + \bar{\lambda} \right] - 2(\bar{H}-\bar{C}) \left[ \bar{H}^2 + (\bar{H} - \bar{r}^{-1} \sin \psi )^2 \right] \;,
\end{array}
\end{equation}
with the boundary conditions
\begin{equation} \label{eq:Snd_bc_axi}
\begin{array}{l}
\bar{r}(0) = \bar{r}_n \;,\\
\bar{z}(0) = 0 \;,\\
\psi(0) = \pi/2 \;,
\end{array} \quad
\begin{array}{l}
\bar{r}(0.5) = \bar{r}_n \cosh [\bar{z}(0.5) /\bar{r}_n] \;, \\
\bar{z}(0.5) = \bar{r}_n \text{asinh} [0.5/\bar{r}_n] \;, \\
\psi(0.5) = \text{asin} [ 1/\sqrt{1+(0.5/\bar{r}_n)^2} ] \;,
\end{array} 
\end{equation}
where $\bar{r}_n = r_n/L$.

\subsection{Numerical implementation and results}

With the aim to verify that the distribution of spontaneous curvature obtained through the inverse problem produces a catenoid-shaped membrane, we chose a different numerical method for the direct computation. Therefore, we solved the system with a custom made code in Matlab\textsuperscript{\tiny\textregistered} (Mathworks, Natick, MA), utilizing the built-in boundary value problem solver \textit{bvp4c}. The values for the distribution of spontaneous curvature was extracted from the data shown in Fig.~3(c) using a spline interpolation to obtain the value of $C$ between the initial mesh-points. Then the system of Eqs.~\ref{eq:Snd_pde_axi} with the boundary conditions \ref{eq:Snd_bc_axi} was solved on an initial mesh of  $1,000$ equidistant points. To obtain convergence, the solver was allowed to increase the number of mesh-points up to $100,000$. The relative and absolute tolerances were set to $10^{-4}$ and $10^{-7}$ respectively.

Solution for the direct computations of the shape equation are presented in Fig.~\ref{fig:direct}. Both the shape and the Gaussian curvature fit closely the ones of a perfect catenoid. This results confirm that the distribution of spontaneous curvature obtained through the inverse approach minimizes the energy of a catenoid-shaped structure subject to the the above boundary conditions.

\end{document}